\documentclass[draft]{agujournal2019}
\usepackage{url} 
\usepackage{lineno}
\usepackage{amsmath}
\usepackage[finalnew]{trackchanges} 
\usepackage{soul}
\nolinenumbers
\draftfalse

\journalname{JGR: Atmospheres}

\begin{document}

%
%


\title{Relativistic runaway electron avalanches within complex thunderstorm electric field structures}

%
%




\authors{E. Stadnichuk\affil{1,2}, E. Svechnikova\affil{4}, A. Nozik\affil{1,5}, D. Zemlianskaya\affil{1,3}, T. Khamitov\affil{1,3}, M. Zelenyy\affil{1,3}, and M. Dolgonosov\affil{6}}


\affiliation{1}{Moscow Institute of Physics and Technology - 1 “A” Kerchenskaya st., Moscow, 117303, Russian Federation}
\affiliation{2}{HSE University  - 20 Myasnitskaya ulitsa, Moscow 101000 Russia}
\affiliation{3}{Institute for Nuclear Research of RAS - prospekt 60-letiya Oktyabrya 7a, Moscow 117312}
\affiliation{4}{Institute of Applied Physics of RAS - 46 Ul'yanov str., 603950, Nizhny Novgorod, Russia}
\affiliation{5}{JetBrains Research - St. Petersburg, st. Kantemirovskaya, 2, 194100}
\affiliation{6}{Space Research Institute of RAS, 117997, Moscow, st. Profsoyuznaya 84/32}




\correspondingauthor{Egor Stadnichuk}{yegor.stadnichuk@phystech.edu}




\begin{keypoints}
\item A new feedback mechanism, reactor feedback, in the dynamics of relativistic runaway electron avalanches is proposed
\item Necessary conditions on the electric field structure for reactor feedback are found
\item Required electric field values are shown to be more achievable for reactor feedback than for relativistic feedback
\end{keypoints}

%
%

%
%


\begin{abstract}
Relativistic runaway electron avalanches (RREAs) are generally accepted as a source of thunderstorms gamma-ray radiation. Avalanches can multiply in the electric field via the relativistic feedback mechanism based on processes with gamma-rays and positrons. This paper shows that a non-uniform electric field geometry can lead to the new RREAs multiplication mechanism --- ``reactor feedback'', due to the exchange of high-energy particles between different accelerating regions within a thundercloud.
A new method for the numerical simulation of RREA dynamics within heterogeneous electric field structures is proposed.
Under the assumption that reactor feedback can produce a terrestrial gamma-ray flash (TGF), necessary conditions for TGF occurrence in the system with the reactor feedback are derived using the developed analytical description and the numerical simulation.
Observable properties of TGFs influenced by the proposed mechanism are discussed.
\end{abstract}

\section{Introduction}

High-energy radiation originating from thunderclouds can be registered by detectors on satellites and on the ground surface. Intense bursts of photons with energy 10~keV -- 100~MeV lasting 0.1--5~ms are called terrestrial gamma-ray flashes (TGFs) and are usually observed from satellites \cite{Fishman1994, Fermi_2016, 10_month_ASIM}. Thunderstorm ground enhancements (TGEs) and gamma-ray glows can be observed under thunderclouds and have a duration of up to several hours \cite{Chilingarian_2011_natural_accelerator, Gurevich2016_tien_shan, Torii2009}. The gamma-radiation of thunderclouds is caused by bremsstrahlung of runaway electrons, which accelerate and multiply in the electric field, forming relativistic runaway electron avalanches (RREAs) \cite{Gurevich1992,Lehtinen1999,Dwyer2012_phenomena}. 
Numerical estimations show that $10^4$--$10^{13}$ RREAs, about $10^6$ runaway electrons in each one, are required to cause a TGF observable from space \cite{doi:10.1002/jgra.50188, Gurevich_2001, Khamiton2020}.
There are two models of TGF production discussed nowadays. The lightning leader model assumes that avalanches emitting gamma-rays originate from thermal electrons accelerated in the strong local electric field of the lightning leader tip \cite{https://doi.org/10.1029/2005JA011350}. The relativistic feedback discharge model \cite{Dwyer_2003_fundamental_limit} considers the multiplication of avalanches which can lead to the self-sustaining development of RREAs: generation of a large number of avalanches even without an external source of high-energy particles \cite{Dwyer2007}.

The relativistic feedback model describes the creation of new avalanches by positrons or energetic photons of the initial avalanche in the region with the electric field strength that exceeds the value required to form RREAs. A new avalanche can be created by a high-energy particle that moves towards the start of an initial avalanche. It should be noted that all the particles, including gamma rays, are radiated mainly along with the avalanche development. Thus, the efficiency of the relativistic feedback mechanism is limited by the probability for a positron or gamma-ray to reverse in the direction opposite to the movement of the avalanche \cite{Dwyer_2012}. The efficiency of the creation of new avalanches can be higher if it does not require a reversal of particle movement. An opportunity for more efficient RREA development in the geometry with two or more avalanche regions with opposite electric field vectors was mentioned in \protect\cite{Dwyer2007}. It was shown in \protect\cite{Kutsyk2011606} that relativistic feedback within a cylindrical electric field caused by a lightning leader is more effective than in a single uniform accelerating region. In this paper, we consider a more general case of the structure with several avalanche regions.

Consider the electric field structure, which is nonuniform on a scale greater than the relativistic runaway electron avalanche growth length. In this case, particles emitted by the initial avalanche can reach regions with the direction of the electric field different from that in the region of the initial avalanche. Thus, the change of direction required for a particle to create a RREA will be smaller than that in the uniform field. For this reason, the initial avalanche can create more new avalanches. Moreover, each of the new avalanches emits particles mainly along with itself, and some of them can reach the region of the initial avalanche, enhancing it. The described processes lead to the creation of new avalanches and amplification of the initial one, and the concept hereinafter is referred to as the ``reactor model'', Figure~\ref{reactor_model}. The new kind of feedback occurring in the nonuniform electric field is called ``reactor feedback''.

This paper presents the numerical simulation and the analytical description of the reactor model. In Section \protect\ref{Simulation} (Simulation) the reactor model is studied via macroscopic Monte-Carlo simulation.
The analytical approach for reactor model study is described in the section Analytical model of multicell reactor structure, Section~\protect\ref{Analytical completely random reactor model}. The spatial distribution and the time dependence of gamma-ray flux are calculated. The conditions for TGF occurrence within a reactor thundercloud are derived. In the Section \ref{Discussion} (Discussion) predictions of the model are compared with observation data and conclusions of other modeling studies. The question of the electric cloud structure and applicability of the model of the reactor structure is addressed. 

\section{Multicell reactor model}
\begin{figure}
    \centering
    \includegraphics[width=1\linewidth]{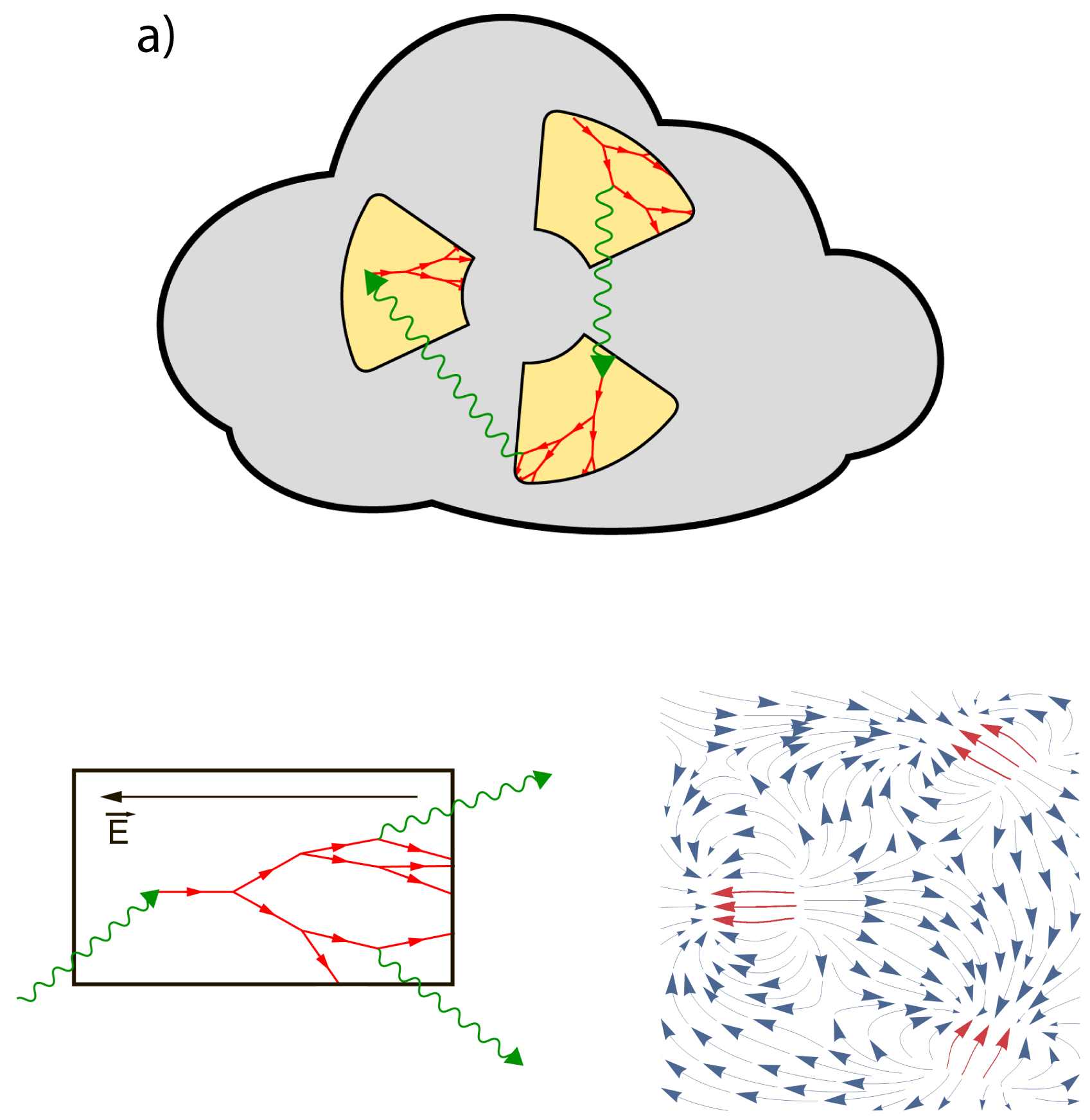}
    \caption{The illustration of the reactor model concept.
    a) The dynamics of relativistic runaway electron avalanches in complex thunderstorm electric field structures. Gamma-ray exchange between electric field regions with electric field strength sufficient for relativistic runaway electron avalanches formation leads to RREA multiplication, which can make avalanches self-sustainable.
    b) The diagram of gamma-ray interaction with an electric field region with electric field strength sufficient for RREA formation. The relativistic runaway electron avalanche created by a gamma-ray produces new gamma rays in this strong electric field region, which results in gamma-ray multiplication.
    c) The distribution of the electric field in the structure described by reactor model \protect\ref{reactor_model}. Regions with red arrows are ``cells'' with the quasi-uniform electric field sufficient for the RREA development. The electric field outside cells is subcritical.
    }
    \label{reactor_model}
\end{figure}

The reactor model describes the interaction of avalanches developing in regions of the strong electric field sufficient for RREA acceleration, which are further called ``cells''. The field strength between cells has under critical value. The ``reactor feedback'' can occur in a thundercloud with a complex electric structure, consisting of several cells with different directions of the electric field \cite{doi:10.1063/1.5130111,EGU}.
Figure~\ref{reactor_model} a) illustrates the interaction of cells in the reactor structure. Let a seed electron form a RREA within one of the cells. The RREA produces gamma-rays via bremsstrahlung. On thunderstorm altitudes, the mean free path of gamma-rays is about several hundred meters or more (400~m for 1~MeV gamma on 10~km altitude \cite{NIST}), so gamma rays can propagate through regions with the subcritical field, and reach other cells and produce RREAs. New RREAs, similarly to the initial one, radiate gamma-rays, which can generate RREAs in other cells of the thundercloud. 
The closer the direction of the field is to the direction to the other cell, the greater the probability of creating a new avalanche in the initial cell by the radiation of secondary avalanches.
By the described way, the complexity of the electric field structure can lead to self-sustainable RREA multiplication due to the exchange of high-energy particles between cells. In other words, RREA in different strong field regions can amplify each other.
A great number of RREAs developed under the influence of the reactor feedback can be sufficient for the production of TGF \cite{doi:10.1063/1.5130111}.





Figure \ref{reactor_model} b) shows the diagram of gamma-ray multiplication in the strong field region. High energy photon interacts with air via Compton scattering, photo-effect, or electron-positron pair production, leading to the production of the high energy electron, which might produce a RREA. The RREA emits gamma-rays, leading to the multiplication of the initial high-energy photon. The electric field outside the cell is subcritical, so the energetic electrons are quickly absorbed by the air. Still, if an energetic electron reaches another cell, it can initiate a RREA, similarly to a gamma-ray.

The reactor feedback can be conveniently discussed within the electric field of the structure hereinafter called ``multicell reactor'' \cite{doi:10.1063/1.5130111}, which consists of a huge number of cells with different directions of the electric field, Figure \ref{reactor_model} c). This structure is an extreme case of electric field geometry, an opposite to another extreme case: a thunderstorm with a uniform electric field. The multicell structure exhibits a chain reaction of gamma-ray interactions with cells. The described high-energy particle dynamics brings to mind the behavior of neutrons in a nuclear reactor. For this reason, the concept of the exchange of relativistic particles between strong field regions is called the ``reactor model''.

The proposed model is further investigated using two approaches: analytical (Section \ref{Analytical completely random reactor model}) and numerical (Section \ref{Macroscopic_simulation}). A macroscopic numerical model provides the ability to take into account the fact that there is a nonzero distance between the point of interaction of initial gamma-ray and the point of production of a new gamma-ray. Among the advantages of the analytical model are a simple account of gamma-ray diffusion and the possibility of explicit consideration of the system parameters required for infinite feedback. The application of the two methods gives independent assessments of the dynamics of the system, while the agreement of the conclusions indicates the reliability of the results.

\section{Simulation}
\label{Simulation}

The movement of runaway electrons is defined by the electric field, while bremsstrahlung gamma-rays can move through the cloud uninfluenced by the electrical structure.
Consequently, RREAs dynamics within a thundercloud can be described as RREAs developing in a region with a strong quasi-uniform electric field and energetic particles propagating between strong field regions and initiating RREAs in it.
For this reason, the behavior of RREAs in the complex electric field structure can be conveniently modeled in two stages: microscopic (RREA development in strong field regions, simulated, for example, using GEANT4 \protect\cite{geant4}) and macroscopic (propagation of particles between regions of RREA development, described by the model developed in this paper and presented in Section \protect\ref{Macroscopic_simulation} (Macroscopic simulation)). In general, the microscopic simulation studies gamma-ray (or other high-energy particles) interaction with cells (Figure~\protect\ref{reactor_model} b)) for various parameters of gamma-rays (energy, momentum direction) and cells (point of interaction, electric field value, cell geometry). The macroscopic simulation calculates the transport of gamma-rays (and possibly other high-energy particles) between cells and raffles out high-energy particles' interactions with cells according to distributions obtained from the microscopic simulation. The approach presented below requires rather less computational time than straightforward modeling.

\subsection{Microscopic simulation}

\begin{figure}[h!]
    \centering
    \includegraphics[width=1.0\linewidth]{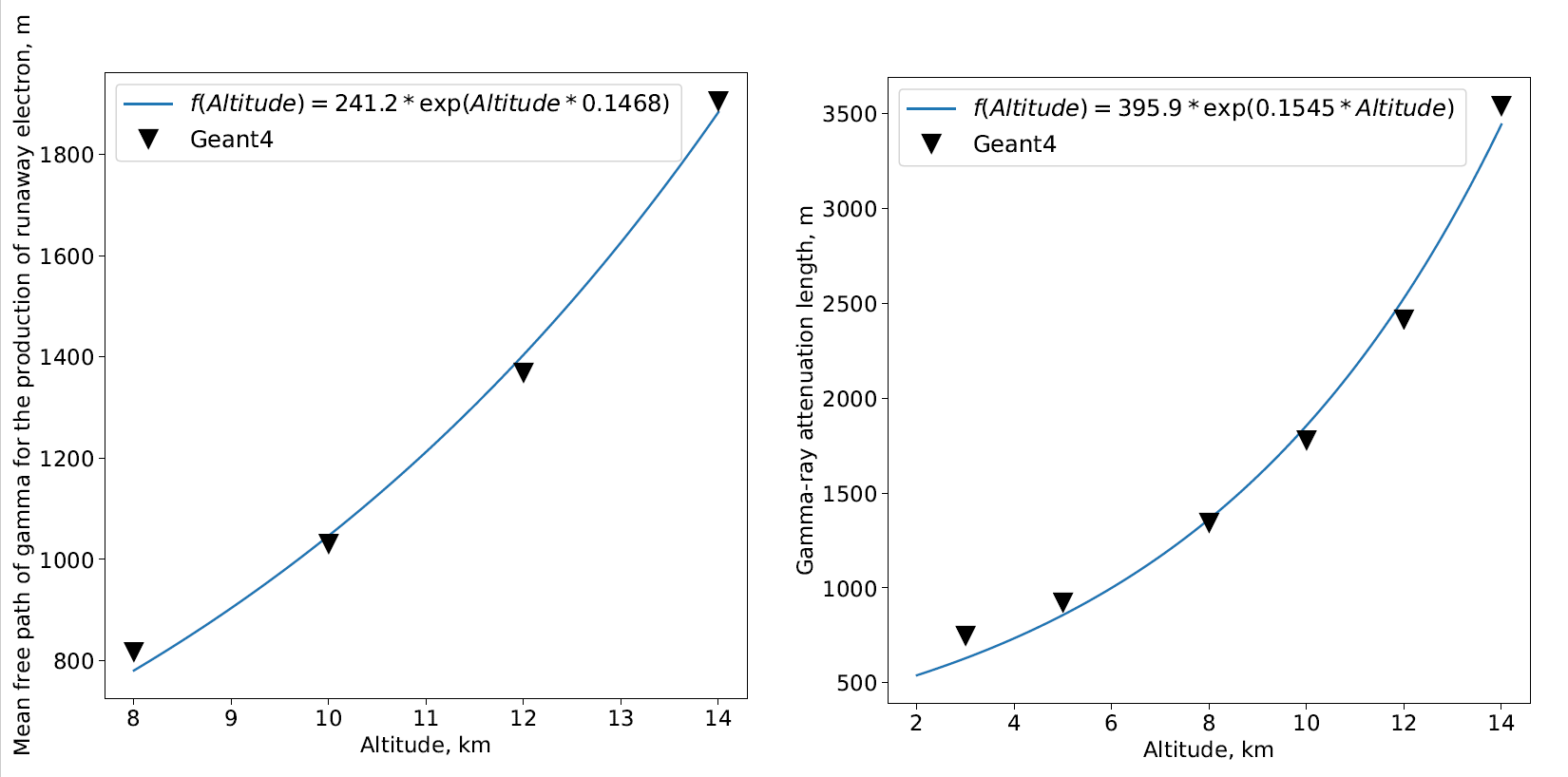}
    \caption{The results of gamma-rays modeling using GEANT4 (black triangles) and approximation (blue curves). Gamma-ray starting energy is set to 7~MeV. Characteristic mean free path of gamma for the production of runaway electron, depending on the atmosphere altitude is on the (left) figure and characteristic gamma-ray attenuation length depending on the altitude is on the (right) figure. The runaway electron production length does not depend on the field strength in the range 1.4--4.3$E_{critical}$, where $E_{critical}$ --- minimum electric field strength required for RREA formation.}
    \label{gamma_decay}
\end{figure}

In this paper, the microscopic simulation was used to calculate characteristic lengths of gamma-ray interaction with air and to calculate the probability of a RREA development after gamma-ray interaction.
Figure~\ref{gamma_decay} presents the dependence of gamma-ray attenuation length and the mean free path before the production of runaway electrons on the thunderstorm altitude, obtained by GEANT4 simulation \protect\cite{geant4}. Gamma-ray attenuation length is a characteristic distance of gamma-ray propagation before it is absorbed by the air. The mean free path before the production of runaway electrons is a characteristic distance traveled by a gamma-ray before runaway electron production via interaction with air molecules (photo-effect, Compton scattering, and electron-positron pair production). It turned out that the vast majority of the electrons produced by gamma-rays are runaway electrons for the electric field strength in the range 1.4--4.3$E_{critical}$, where $E_{critical}$ is the minimum electric field strength required for RREA formation. For this reason, the dependence of the mean free path of gamma for the production of runaway electrons on the electric field is negligible. Also, since gamma does not interact with the electric field directly, gamma-ray attenuation length does not depend on the electric field strength.

A high-energy particle interacting with a cell produces a seed electron that can initiate a RREA. The momentum direction of a generated seed electron is random for the multicell reactor, thus, in general, this electron has to turn in the direction against the cell electric field to produce a RREA. Consequently, one of the crucial parameters of the system is the probability of a reversal of the generated electron with RREA development. In this paper, the GEANT4 simulation was carried out to calculate the reversal probability depending on the parameters of the electric field structure. In the present simulation, seed electrons were launched from the middle of the cell. The resulting RREA was registered using the detector modeled at the edge of the cell. If the seed electron produces a RREA then it has reversed, otherwise, it was absorbed and it did not have any further impact on RREAs dynamics within the thundercloud. In this study, the electron reversal probability was calculated as the number of reversed seed electrons divided by the total number of launched seed electrons.

\begin{figure}[h!]
\begin{minipage}{0.5\linewidth}
    \centering
    \includegraphics[width=1.1\linewidth]{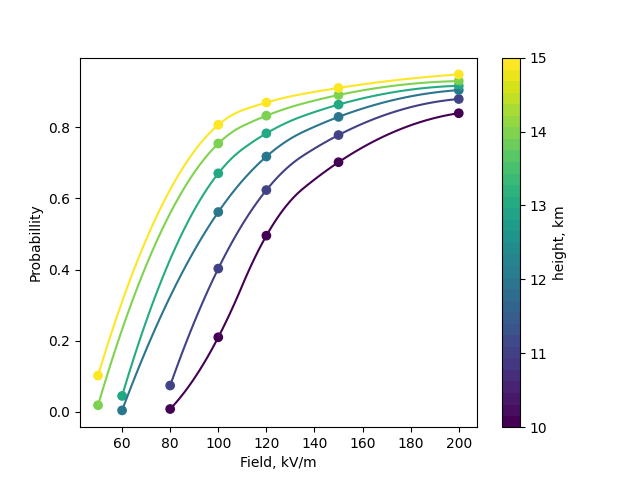}
    \caption{Probability for a 1 MeV seed electron with random momentum direction to produce a RREA depending on the electric field strength: the GEANT4 modeling results (dots) and the quadratic interpolation (lines).}
    \label{reversal_probability}
\end{minipage}
~
\begin{minipage}{0.5\linewidth}
    \centering
    \includegraphics[width=1.1\linewidth]{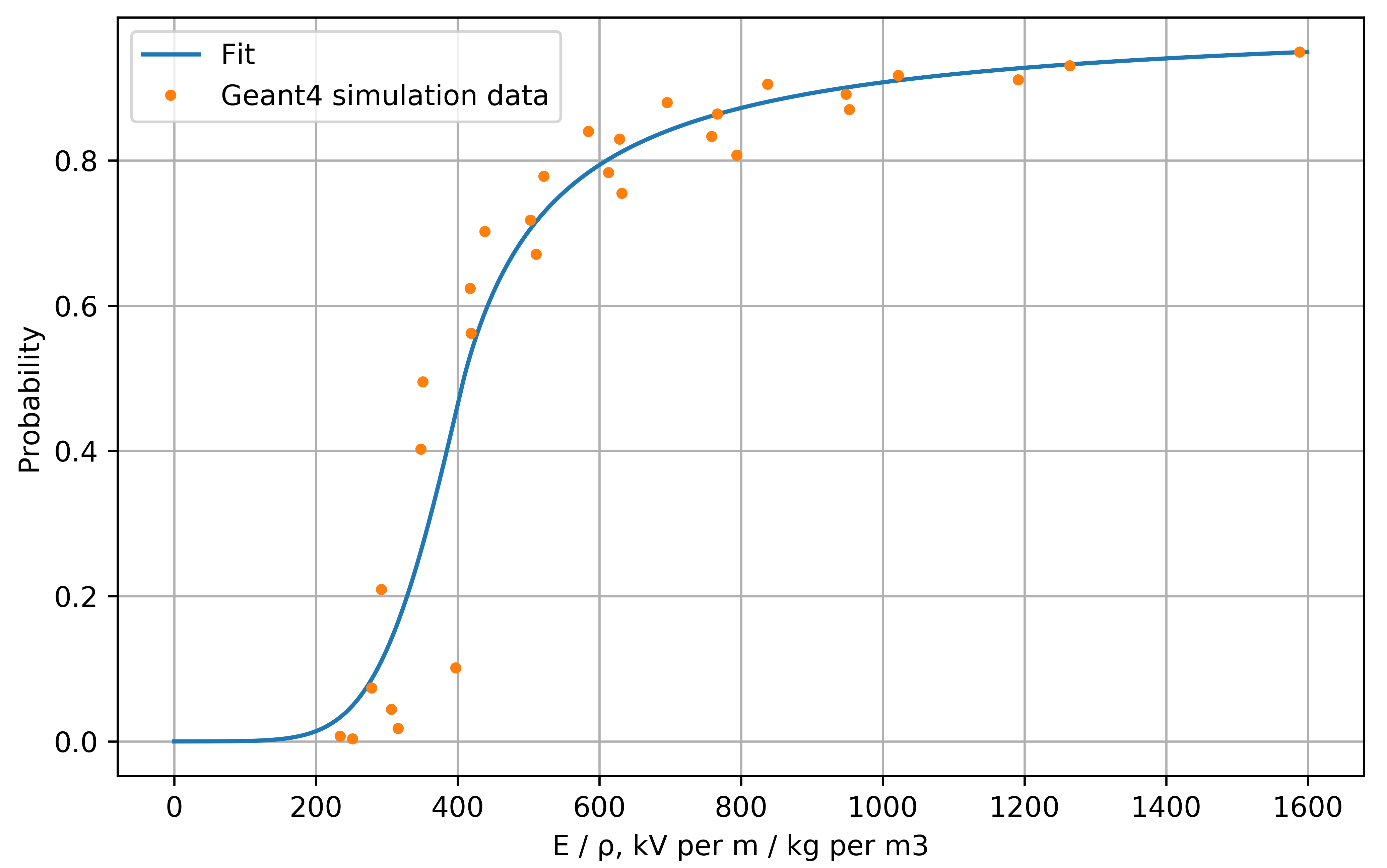}
    \caption{Probability for a 1 MeV seed electron with random momentum direction to produce a RREA depending on the ratio of the electric field strength and the air density ($\frac{E}{\rho}$): the fit with sigmoid function \ref{fit_formula} for the GEANT4 simulation for the 1 MeV electron.}
    \label{reversal_fit}
\end{minipage}
\end{figure}

In general, the electron reversal probability depends on the electric field strength in the cell, air density, seed electron energy, and momentum direction. In this paper, the probability of a seed electron to produce a RREA was modeled for an isotropic source of 1~MeV (characteristic runaway electron energy) seed electrons, Figure~\ref{reversal_probability},\ref{reversal_fit}. In the case of the subcritical electric field, RREAs can not develop, which means that the probability of RREA generation is 0. For the electric field higher than the critical value the probability is close to 1. The characteristic spatial scale of electron reversal appeared to be below 2~m for 1~MeV electron, which is much less than the typical size of the cell, therefore, the electron reversal spatial scale is negligible. Let $P$ be electron reversal with RREA formation probability, $E$ --- electric field strength in kV/m, $\rho$ --- air density, kg/m${}^{3}$. Hence, simulation results can be fit with the empirical relation, Formula~\protect\ref{fit_formula}:

\begin{equation}
    P\left( \frac{E}{\rho} \right) = 
    \begin{cases}
        \frac{1}{2} \cdot \left( 1 + erf \left(3.0378 \frac{E}{\rho} - 0.0074 \right) \right) &\text{, $3.0378 \frac{E}{\rho} - 0.0074 \le 0$}\\
        \frac{1}{2} \cdot \left(1 + \frac{3.0378 \frac{E}{\rho} - 0.0074}{1 + \left(3.0378 \frac{E}{\rho} - 0.0074 \right)} \right) &\text{, $3.0378 \frac{E}{\rho} - 0.0074 \ge 0$}
    \end{cases}
    \label{fit_formula}
\end{equation}

\subsection{Macroscopic simulation}
\label{Macroscopic_simulation}
Contrary to the microscopic simulation carried out using GEANT4, the macroscopic modeling operates with averaged parameters of energetic particles and does not take into account individual events of particle interaction.
The interaction of cells is caused mainly by high-energy photons because their movement is not influenced by the electric field and the interaction with air is rather smaller than that for electrons. The impact of the runaway electron transport between cells can be neglected. For this reason, the performed macroscopic modeling characterizes the propagation of high-energy photons between strong field regions within the thundercloud.

The macroscopic model is implemented in Kotlin \cite{Kotlin}. Simulation describes two types of particles: runaway electrons (with energy higher than the critical value for given altitude and electric field) and photons (with the energy above the minimum energy of runaway electrons) capable of creating runaway electrons via interactions with air molecules. Each particle is characterized by the origin point. The movement of the particle is described by the velocity vector and energy.

Within a macroscopic simulation, cells can be implemented in two different ways. The first way is to divide the thundercloud volume into cells before the simulation run. The second way is to generate cells on the run: in this case, the start of the cell is defined as the point of a RREA production. The second option is implemented in the modeling described below.

The macroscopic simulation is based on the following assumptions:
\begin{itemize}
    \item A photon moves in the same direction until the interaction. Distance between the origin point and the interaction point is described by the exponential dependence with mean free path calculated in microscopic modeling, Figure~\ref{gamma_decay}.
    \item A photon produces the electron with the same energy and direction. The angular distribution of produced electrons is considered indirectly via electron reversal probability, Figure~\protect\ref{reversal_fit}. Our calculations show that the applied approach and direct consideration of angular distribution lead to similar results.
    \item The field direction in the point of the electron production is chosen, assuming that the probability of each field direction is equal.
    \item The structure consists of a large number of cells ($\geq 10$).
    \item Bremsstrahlung photons of the RREA are generated at a fixed distance (the avalanche length) in the direction of the electric field at the point of the RREA origin. All generated photons have the same energy and move alongside the electric field in the RREA origin. The number of bremsstrahlung photons generated by the RREA follows the Poisson distribution with a given average which is called the local multiplication factor.
\end{itemize}

The multiplication factor is the ratio of the number of particles in one generation to that of the previous generation.
The local multiplication factor is the mean number of gamma-rays generated by one gamma ray in one multiplication process, Figure \ref{reactor_model} b). The described simplifications give the model an important advantage: the opportunity to characterize the system dynamics using only two parameters --- the size of the modeling region (the size of the cloud, which is considered cubic) and the local multiplication factor. The avalanche length has a small effect on the simulation results. The local multiplication factor depends on many parameters including the angle between the electron velocity and the electric field in the strong field region (for large angles, the electron most probably ``dies'' without starting the avalanche) and the actual distribution of the field inside the cell. A more detailed discussion of the local multiplication factor is presented in Section \protect\ref{local_multiplication_factor} (Local multiplication factor).

\begin{figure}[h!] 
    \centering
    \includegraphics[width=1.0\linewidth]{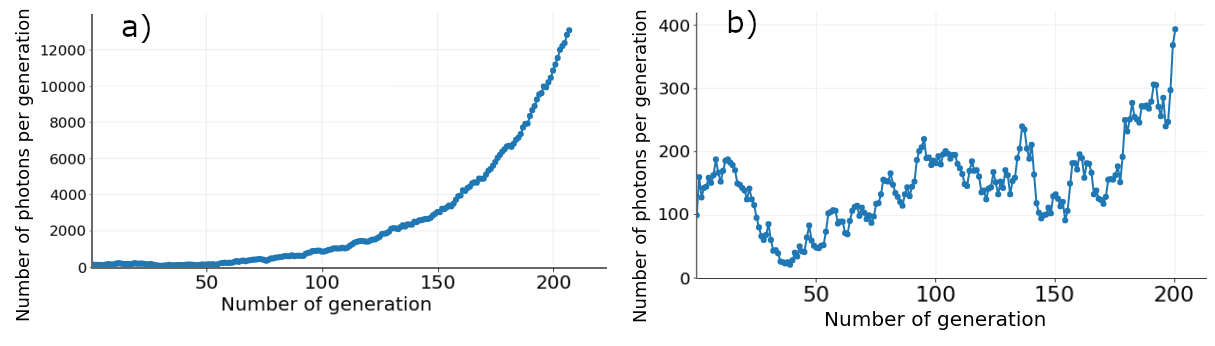}
    \caption{The dependence of the number of gamma-rays on the gamma-ray generation number, calculated using the macroscopic simulation. The average lifetime of one generation is approximately one microsecond. Cell length is set to 300~m, mean free path of photons is set to 100~m, the initial number of high energy photons is 100. 
    (a) a TGF-like mode with the rapid increase of the number of gamma-rays: multiplication factor is 1.5 (cloud size is 1250~meters). 
    (b) a mode similar to a gamma-ray glow or TGE: a long-duration flux of approximately constant intensity. The multiplication factor is close to 1 (cloud size is 1200~m).
}
    \label{macroreactor}
\end{figure}

Figure \ref{macroreactor} illustrates the rate of production of high-energy photons in the multicell reactor model. The number of gamma-rays depends on the gamma generation number. The concept of the gamma-ray generations works as follows. At the start of a macroscopic simulation, initial gamma-rays are launched. This is the first generation of reactor feedback. These gamma rays propagate through the thunderstorm media and eventually interact with cells. Interactions with cells produce new gamma-rays of the next generation. These gamma-rays produce gamma of further generations via their interactions with cells. Thus, the dependence of the number of gamma-rays on their generation number shows how reactor feedback operates in the multicell reactor thundercloud system. A lifetime of one generation is the meantime of photon propagation before its interaction, it can be estimated as cell length plus gamma-ray free path length divided by the speed of light, which gives about 1 $\mathrm{\mu s}$. In more detail, the issue of the relationship between the generation number and time is discussed in Section \ref{comparison} (Comparison of theory and simulation). Figure \ref{macroreactor}(a) demonstrates the dramatic increase of the number of gamma rays on the time scale of TGF. Figure \ref{macroreactor}(b) is obtained for the electric field structure of smaller size (1200~m instead of 1250~m), which leads to a decrease of multiplication factor down to~1. As a result, the system exhibits a TGE-like mode with approximately constant energetic particle flux.

\section{Analytical model of multicell reactor structure}
\label{Analytical completely random reactor model}

The developed analytical model of the gamma-ray dynamics in the multicell reactor is based on assumptions similar to the macroscopic simulation. Nevertheless, there are some differences between the theory and the macroscopic simulation. The assumptions of the theoretical model are listed below:

\begin{itemize}
    \item The electric field direction is arbitrary in each strong field region, which makes gamma-ray local multiplication isotropic.
    \item The electric field outside cells is subcritical.
    \item The critical electric field and the air density have the same value in all cells.
    \item The structure consists of a large number of cells ($\geq 10$).
    \item All gamma-rays have the same energy determined by the bremsstrahlung of RREAs.
    \item Gamma-ray emitted by the RREA is generated in the point of interaction of the initial gamma-ray leading to the production of this RREA, Figure \ref{reactor_model} b).
    \item The energetic photon can leave the system in two ways: by escaping the thundercloud or by losing energy via the production of a runaway electron.
    \item The system is axially symmetrical. The reactor is a cylinder with a height H and a radius R.
\end{itemize}

With the assumptions above, the dynamics of gamma-rays in the thundercloud can be described by the reactor diffusion equation:

\begin{equation}
    D \Delta n(t, r, z) - c \Sigma n(t, r, z) + \nu c \Sigma n(t, r, z) = \frac{\partial n(t, r, z)}{\partial t}
    \label{reactor_equation}
\end{equation}

This equation describes gamma-ray multiplication by interaction with cells (Figure \protect\ref{reactor_model} b)) as a chain reaction (Figure \ref{reactor_model}). $n(\vec{r},t,z)$ is gamma-ray concentration, $D = \frac{c \lambda}{3}$ --- diffusion coefficient, $\lambda$ --- mean free path length for gamma-rays, $\Sigma = \frac{1}{\lambda_{\gamma \rightarrow e^-}}$ --- mean macroscopic cross-section of runaway electron production by a gamma ray, $\nu$ --- local multiplication factor, and $c$ --- speed of light. The mentioned parameters are defined by the structure of the electric field and by air density.

The term $-c \Sigma n$ is responsible for gamma-ray extinction via the production of runaway electrons. The term $\nu c \Sigma$ is responsible for gamma-ray production via RREA bremsstrahlung (Section \ref{local_multiplication_factor}, Local multiplication factor). The creation of a RREA takes a considerable amount of energy from the photon, which is absorbed shortly afterward. For this reason, we use the assumption that macroscopic cross-sections of gamma extinction and gamma multiplication are equal, as two parameters describe the same process.

The Laplace operator for the system with the axial symmetry is written as follows:

\begin{equation}
    \Delta_2 + \frac{\partial ^2}{\partial^2 z}
\end{equation}

The outflow of particles from the cloud is described by the following boundary conditions:

\begin{eqnarray}
    n(t, r, z)|_{r = R} = 0,\\
    n(t, r, z)|_{z = 0, H} = 0
\end{eqnarray}

Let us present an eigenfunction as the product of the spatial and the temporal parts:

\begin{equation}
    n(r,z,t) = N_{km}(t) n_{km}(r,z)
\end{equation}

Here k and m are indexes of the reactor equation \protect\ref{reactor_equation} eigenfunctions. Taking into account the boundary conditions, $n_{km}$ are taken as eigenfunctions of the Laplace operator:

\begin{equation}
    n_{km} (r,z) = J_k \left(\frac{a_k r}{R}\right) sin\left(\frac{(m + 1) \pi z}{H} \right)
\end{equation}

Here $a_k$ are zeros of Bessel functions. The temporal part of the solution is described by the following equation:

\begin{equation}
    N_{km}(t) \bigg( \frac{3 (\nu - 1)}{\lambda \lambda_{\gamma \rightarrow e^-}} - \left(\frac{a_k}{R}\right)^2 - \left(\frac{(m + 1)\pi}{H}\right)^2 \bigg) = \frac{3}{\lambda c} \frac{d N_{km}}{d t}
\end{equation}

For simplicity, the initial condition is chosen as follows:

\begin{equation}
    N_{km}|_{t = 0} = N_0 = const,
\end{equation}

which leads to the following solution:

\begin{eqnarray}
    n(r,z,t) = N_0 \cdot \sum^{\inf}_{k, m = 0} J_k\left(\frac{a_k r}{R}\right) sin\left(\frac{(m + 1) \pi z}{h}\right) e^{\varepsilon_{km} t},\\
    \varepsilon_{km} = \frac{\lambda c}{3} \bigg(\frac{3 (\nu - 1)}{\lambda \lambda_{\gamma \rightarrow e^-}} - \left(\frac{a_k}{R}\right)^2 - \left(\frac{(m + 1)\pi}{H}\right)^2\bigg)
    \label{eq:reactor_spatial}
\end{eqnarray}

Infinite feedback occurs when at least one of the terms in \ref{eq:reactor_spatial} has $\varepsilon_{km} > 0$. The higher k and m, the lower $\varepsilon_{km}$ is. Consequently, if $\varepsilon_{00}$ is slightly more than~0 then other terms decrease over time. Assuming that the multicell reactor becomes discharged earlier than the second term of the sequence starts to grow, only the first term determines the gamma-ray dynamics:

\begin{eqnarray}
    n(r,z,t) = N_0 \cdot J_0\left(\frac{a_0 r}{R}\right) sin\left(\frac{\pi z}{H}\right) e^{\varepsilon t},\\
    \varepsilon = \frac{\lambda c}{3} \bigg(\frac{3 (\nu - 1)}{\lambda \lambda_{\gamma \rightarrow e^-}} - \left(\frac{a_0}{R}\right)^2 - \left(\frac{\pi}{H}\right)^2\bigg)
    \label{theory_formula_global_factor}
\end{eqnarray}

$a_0 \approx 2.405$. $\varepsilon$ is called the ``global multiplication factor'': if $\varepsilon > 0$ then the number of gamma-rays produced by the reactor-like thunderstorm grows exponentially, in other words, the reactor system explodes. Thus, the criterion of the reactor explosion is as follows:

\begin{equation}
    \frac{\lambda c}{3} \bigg(\frac{3 (\nu - 1)}{\lambda \lambda_{\gamma \rightarrow e^-}} - \left(\frac{a_0}{R}\right)^2 - \left(\frac{\pi}{H}\right)^2\bigg) > 0
    \label{explosion_criterion}
\end{equation}

The criterion of reactor explosion not only depends on the local properties of the electrical structure characterized by the local multiplication factor $\nu$. Whether there is a gamma-ray explosion or not depends on the size of the thundercloud as well. The larger the reactor, the smaller the value of the electric field is required for the explosion. It should be noted that for the spatially infinite thundercloud ($R = \infty$, $H = \infty$) the criterion of the explosion takes the form $\nu > 1$.

Gamma-ray flux generated by the multicell reactor thundercloud can be simply derived from the Formula $\Phi = D \nabla n$. As TGFs are observed mostly from the top and the bottom of thunderstorms, we consider the case of observation close to the zenith or nadir, then the flux is as follows:

\begin{equation}
    |\Phi(r,t)|\bigg|_{z=0,H} = \frac{\lambda c}{3}\frac{\partial n(r,z,t)}{\partial z}\bigg|_{z=0,H} = \frac{\pi \lambda c}{3 H} N_0 \cdot J_0\left(\frac{2.405 \cdot r}{R}\right) e^{\varepsilon t}\
    \label{flux_on_the_axis}
\end{equation}

The equation~\ref{flux_on_the_axis} describes the exponential growth of the flux typical of the beginning of TGF and characterizes the dependence of flux on the radius from the axis of the system. In other words, it describes a multicell reactor radiation pattern.

The proposed analytical consideration is designed to describe the multicell structure consisting of a large number of cells. We describe the parameters of the distribution of the electric field by the integral characteristic --- the global multiplication factor. In principle, the global multiplication factor can be defined and used for a structure with a small number of cells, which would require describing the dependence of the global multiplication factor on all details of the structure, including the location of all the cells. However, for a large number of cells, which is assumed for the analytical model, the global multiplication factor depends on the averaged characteristics only (average cell size, the average distance between cells, mean electric field strength, mean air density). Therefore, the analytical model describes the multicell structure using the global multiplication factor depending on the averaged parameters of the structure.

\subsection{Comparison of theory and simulation}
\label{comparison}

\begin{figure}[h!]
    \centering
    \includegraphics[width=1\textwidth]{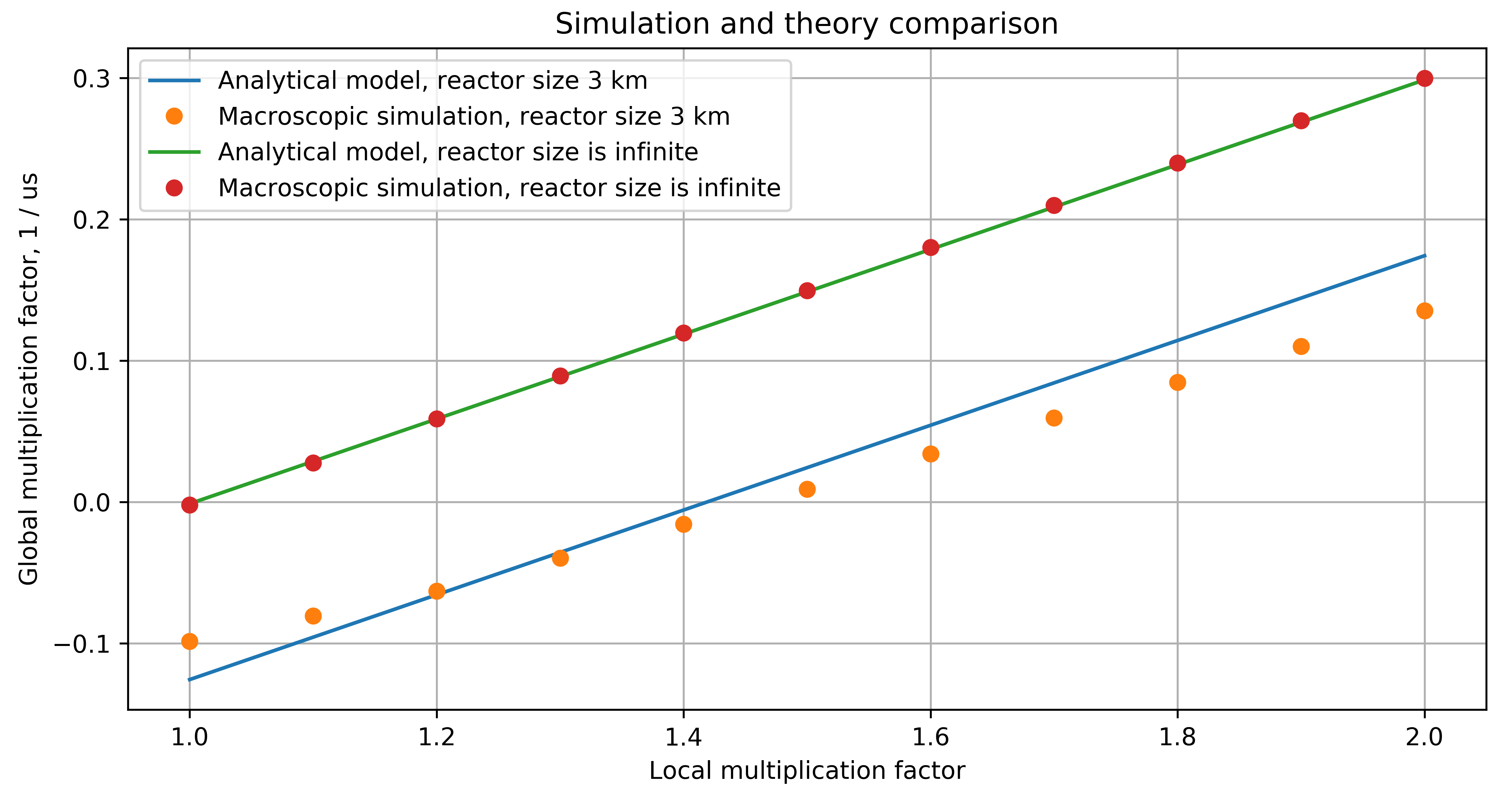}
     \caption{
    The comparison between the macroscopic simulation (Figure~\ref{macroreactor}) and the theoretical model of the multicell reactor (Formula \ref{theory_formula_global_factor}). The graph was plotted for 10~km thunderstorm altitude, where gamma-ray interaction length is equal to approximately 1~km, Figure~\ref{gamma_decay}. When reactor size is much bigger than the characteristic gamma-ray interaction length, the simulation predicts the same RREA dynamics as the theory. Differences for a finite reactor arise due to differences in models assumptions.
}
    \label{models_comparison}
\end{figure}

Both macroscopic simulation and theoretical model show the same properties of the multicell reactor model. Models predict exponential growth of the number of relativistic particles within reactor thunderstorm with an e-folding factor called global multiplication factor, Formula~\protect\ref{theory_formula_global_factor}, Figure~\protect\ref{macroreactor}. Global multiplication factor depends on local multiplication factor, because the stronger the gamma locally multiplies, the faster the total number of relativistic particles grows. Also, both models predict that the global multiplication factor depends on thunderstorm size, because the smaller the reactor is, the easier it is for the gamma to leave it. Still, there are several differences between models. Firstly, the simulation considers the reactor to be cubic and the theory considers it to be a cylinder. Secondly, the macroscopic simulation describes gamma-ray motion between interactions with cells straightforwardly, but the theory considers it to be Brownian. Consequently, it is easier for simulation gamma-rays to leave the thunderstorm, which leads the simulation global multiplication factor to be slightly less than the theoretical one, Figure~\protect\ref{models_comparison}. Thirdly, the macroscopic simulation is able to consider the effect of the length of cells on the geometry of gamma-ray multiplication, when the theoretical model considers local multiplication at the same point as gamma-ray interaction. This leads simulation thunderstorm size to be effectively smaller than the theoretical one, causing simulation global multiplication factor to be smaller. Therefore, for clarity, model comparison in Figure~\protect\ref{models_comparison} was plotted assuming simulation parameter ``cell length'' be equal to zero.

An important question is a duration of one generation within the simulation, Figure~\protect\ref{macroreactor}. Despite the differences between the theoretical model and the simulation, they both show the same behavior for infinite thunderstorms, when thunderstorm size is much higher than gamma-ray interaction length. Within the infinite thunderstorm, when gamma-rays do not leave the multiplication system, the number of simulation gamma-rays grows with generation as $\nu^i$, where $\nu$ --- local multiplication factor, $i$ --- number of generation, and theoretical number of gamma-rays grows with time as $exp\left(\frac{c}{\lambda_{\gamma \rightarrow e^-}}(\nu - 1)t\right)$, according to Formula \protect\ref{theory_formula_global_factor}. Therefore, the duration of one generation within the macroscopic simulation can be estimated as $\frac{\lambda_{\gamma \rightarrow e^-}}{c}\frac{ln\nu}{\nu - 1}$. The reason why one generation duration depends on the local multiplication factor is as follows. It is impossible to unambiguously compare generation and time because in the macroscopic simulation gamma-ray path is raffled out of exponential distribution, which makes the time not fixed. It turns out that different generations can exist at the same time. And the higher the local multiplication factor is, the more confusion between time and generations is introduced. This is taken into account in the generation-time conversion factor $\frac{\lambda_{\gamma \rightarrow e^-}}{c}\frac{ln\nu}{\nu - 1}$, which shows the average duration of one generation.

The figure \protect\ref{models_comparison} shows that both models predict similar global multiplication factors. Global multiplication factor equal to 0.1 $us^{-1}$ means that a reactor thunderstorm will produce $10^{19}$ gamma-rays from a single seed gamma-ray in 450 us. Global multiplication factor being equal to 0.45 $us^{-1}$ produces $10^{19}$ gamma-rays from a single seed gamma in 100 us. Such and even stronger gamma multiplication conditions are hypothetically achievable in the multicell reactor (Figure \protect\ref{local_explosion_criteria}), which means that the proposed reactor model is potentially suitable to describe rise times of terrestrial gamma-ray flashes \protect\cite{10_month_ASIM}.

\subsection{Local multiplication factor}
\label{local_multiplication_factor}

The local multiplication factor is the average number of gamma-ray photons produced by the initial gamma-ray photon as a result of its interaction with a cell, Figure~\protect\ref{reactor_model} b). Let the value of the electric field within a cell be $E$, air density --- $\rho$, cell length --- $L$. These parameters determine the local multiplication factor, which can be described analytically in the following way. Let gamma-ray produce a runaway electron with random momentum direction at the beginning of a cell. Let the probability of the runaway electron reversal with RREA formation be equal to $P$, Formula~\ref{reversal_probability}. This probability includes electron reversal so that it moves in the direction opposite to the electric field direction and RREA formation after reversal. In this study, the probability of reversal is calculated using GEANT4, Figure~\ref{reversal_fit}. $\lambda_{e^- \rightarrow \gamma}$ is the mean path of a runaway electron before production of the energetic photon which can produce a runaway electron avalanche ($\lambda_{e^- \rightarrow \gamma}$ is approximately equal to 500~m for gamma-rays with energy higher than 1~MeV for 10~km atmosphere altitude according to GEANT4 simulations). The RREA e-folding length can be described as follows \cite{Dwyer2007}:

\begin{equation}
    \lambda_{RREA} = \frac{7300 ~keV}{E - \frac{\rho}{\rho_0} \cdot 276 ~\frac{kV}{m}}
    \label{avalanche_length}
\end{equation}

Here $\rho_0$ is the air density at sea level. If a gamma-ray produces a RREA at the beginning of the cell with probability $P$, then the number of gamma-rays radiated by this avalanche can be found from the following equation:

\begin{equation}
    dN_{\gamma}(z) = \frac{dz}{\lambda_{e^- \rightarrow \gamma}} \cdot P \cdot e^{\frac{z}{\lambda_{RREA}}}
\end{equation}

Consequently, the RREA during all the development produces the following number of gamma-rays:

\begin{equation}
    N_{\gamma}(L) = P \cdot \frac{\lambda_{RREA}}{\lambda_{e^- \rightarrow \gamma}} \cdot \left( e^{\frac{L}{\lambda_{RREA}}} - 1 \right)
\end{equation}

In the multicell reactor model, a gamma-ray can interact with air in the cell at any point. Therefore local multiplication factor should be found as a result of the following averaging:

\begin{equation}
    \nu = \int_0^L \frac{dl}{L} N_{\gamma}(l)
\end{equation}

Thus, the local multiplication factor is defined according to Formula~\ref{formula_nu_gamma}:

\begin{equation}
    \nu = \frac{P}{L} \frac{\lambda_{RREA}}{\lambda_{e^- \rightarrow \gamma}}\left(\lambda_{RREA}e^{\frac{L}{\lambda_{RREA}}} - \lambda_{RREA} - L\right)
    \label{formula_nu_gamma}
\end{equation}

Multiplication of gamma-rays takes place in cells, outside cells particles do not multiply, though gamma-rays and electrons can interact with media outside cells. The local multiplication factor characterizes the probability of multiplication, which is dependent on the size of cells and the distance between them. Thus, the local multiplication factor characterizes the geometrical parameters of the cell structure. As a first approximation, the effect of the distance between cells can be considered within the local multiplication factor. Let the characteristic distance between cells be equal to $l$. The probability for a gamma-ray, which was produced in one cell, to reach another cell can be estimated as $exp\left(-\frac{l}{\lambda_-}\right)$. When gamma reaches a cell, it interacts within the cell with a probability that can be estimated as $1 - exp\left(-\frac{L}{\lambda_-}\right)$. In this way, the local multiplication factor can be modified, Formula~\ref{cell_distance_local_multiplication_factor}. The correction that occurs in Formula~\ref{cell_distance_local_multiplication_factor} makes the infinite feedback conditions more strict: cell length required for TGF production for a given electric field strength increases approximately by $l \frac{\lambda_{RREA}}{\lambda_-}$. Therefore, for distances between cells in order of 1~km cell length required for infinite feedback increases by $\lambda_{RREA}$, which is approximately 100~m. It should be mentioned that there is a probability that gamma will not meet strong field regions on its way. For the multicell reactor structure, this probability is rather low because of the large number of cells, so it is neglected.

\begin{equation}
	\nu = e^{-\frac{l}{\lambda_-}} \left(1 - exp\left(-\frac{L}{\lambda_{\gamma \rightarrow e^-}}\right)\right) \frac{P}{L} \frac{\lambda_{RREA}}{\lambda_{e^- \rightarrow \gamma}}\left(\lambda_{RREA} e^{\frac{l}{\lambda_{RREA}}} - \lambda_{RREA} - L\right)
	\label{cell_distance_local_multiplication_factor}
\end{equation}

\subsection{Local multiplication factor with electron transport between cells}

In the previous section, it was assumed that avalanche regions exchange only gamma-rays with each other. In this section, we take into account the exchange of runaway electrons between cells (the impact of the distance between cells on runaway electron transport is discussed in Section \protect\ref{Appendix_D}, Appendix C). Runaway electron transport effectively lengthens relativistic runaway electron avalanches due to their continuation in other cells, thus, it can be taken into account within local multiplication factor as gamma-rays radiated in other cells by transported RREAs. The RREA development in the first cell, which was ignited by a gamma-ray of the previous generation, results in the following number of runaway electrons within this cell:

\begin{equation}
    N_{e^-,1} = P \int_0^L \frac{dl}{L} e^{\frac{l}{\lambda_{RREA}}} = P \frac{\lambda_{RREA}}{L} \left(e^{\frac{L}{\lambda_{RREA}}} - 1\right)
\end{equation}

and gamma-rays:

\begin{equation}
    N_{\gamma,1} = \frac{P}{L} \frac{\lambda_{RREA}}{\lambda_{e^- \rightarrow \gamma}}\left(\lambda_{RREA}e^{\frac{l}{\lambda_{RREA}}} - \lambda_{RREA} - L\right)
\end{equation}

Index 1 means the number of the cell respectively to runaway electron transport between cells. A runaway electron can enter the neighboring cell both along the field and against the field. If the runaway electron enters the cell along the electric field, it decelerates and does not produce gamma-rays. On the contrary, entering the cell against the electric field accelerates the electron, allowing the RREA creation. In the case of the multicell reactor, the probability of the electron acceleration in the cell is $0.5$. Consider the probability of a runaway electron transport between cells is $\delta$. Therefore, on average, $0.5 \cdot \delta$ of transported runaway electrons form a new avalanche, which influences the local multiplication factor as follows. Runaway electrons reaching another (second) cell can form RREAs at the beginning of the second cell. That leads to the following number of runaway electrons at the end of the second cell:

\begin{equation}
    N_{e^-,2} = 0.5\delta P \frac{\lambda_{RREA}}{L} \left(e^{\frac{L}{\lambda_{RREA}}} - 1\right) \cdot e^{\frac{L}{\lambda_{RREA}}}
\end{equation}

RREAs developed in the second cell radiate gamma-rays:

\begin{equation}
    N_{\gamma,2} = 0.5\delta P \frac{\lambda_{RREA}}{L} \left(e^{\frac{L}{\lambda_{RREA}}} - 1\right) \cdot \frac{\lambda_{RREA}}{\lambda_{e^- \rightarrow \gamma}}\left(e^{\frac{L}{\lambda_{RREA}}} - 1\right)
\end{equation}

Similarly, the number of energetic particles in the third cell will differ from that in the second cell by the factor $0.5 \delta e^{\frac{L}{\lambda_{RREA}}}$. Therefore, the local multiplication factor influenced by runaway electron transport can be calculated as follows:

\begin{equation}
    \nu = N_{\gamma} + N_{\gamma,2} \cdot \sum_0^{+\infty}0.5\delta e^{\frac{L}{\lambda_{RREA}}}
\end{equation}

This results in the following formula for local multiplication factor with runaway electron transport between cells:

\begin{equation}
\begin{split}
    \nu & = \frac{P}{L} \frac{\lambda_{RREA}}{\lambda_{e^- \rightarrow \gamma}}\left(\lambda_{RREA}e^{\frac{l}{\lambda_{RREA}}} - \lambda_{RREA} - L\right) +  0.5\delta P \frac{\lambda_{RREA}}{L} \cdot \\
    & \cdot \left(e^{\frac{L}{\lambda_{RREA}}} - 1\right) \frac{\lambda_{RREA}}{\lambda_{e^- \rightarrow \gamma}}\left(e^{\frac{L}{\lambda_{RREA}}} - 1\right) \cdot \sum_0^{+\infty}0.5\delta e^{\frac{L}{\lambda_{RREA}}}
\end{split}
\end{equation}

To consider a finite thundercloud we should limit the number of terms in the sum to $\approx \frac{L}{R}$, where $R$ is a characteristic size of the thunderstorm. In what follows, for simplicity, the infinite sum is calculated. For the case $0.5\delta e^{\frac{L}{\lambda_{RREA}}} < 1$ the local multiplication factor gets the following form:

\begin{equation}
    \begin{split}
        \nu & = \frac{P}{L} \frac{\lambda_{RREA}}{\lambda_{e^- \rightarrow \gamma}}\left(\lambda_{RREA}e^{\frac{l}{\lambda_{RREA}}} - \lambda_{RREA} - L\right) + \\
        & + \frac{\lambda_{RREA}}{\lambda_{e^- \rightarrow \gamma}}\left(e^{\frac{L}{\lambda_{RREA}}} - 1\right) \cdot \frac{0.5\delta P \frac{\lambda_{RREA}}{L} \left(e^{\frac{L}{\lambda_{RREA}}} - 1\right) }{1 - 0.5\delta e^{\frac{L}{\lambda_{RREA}}}}
    \end{split}
    \label{formula_nu_rreas}
\end{equation}

\begin{figure}[h!]
    \centering
    \includegraphics[width=1\textwidth]{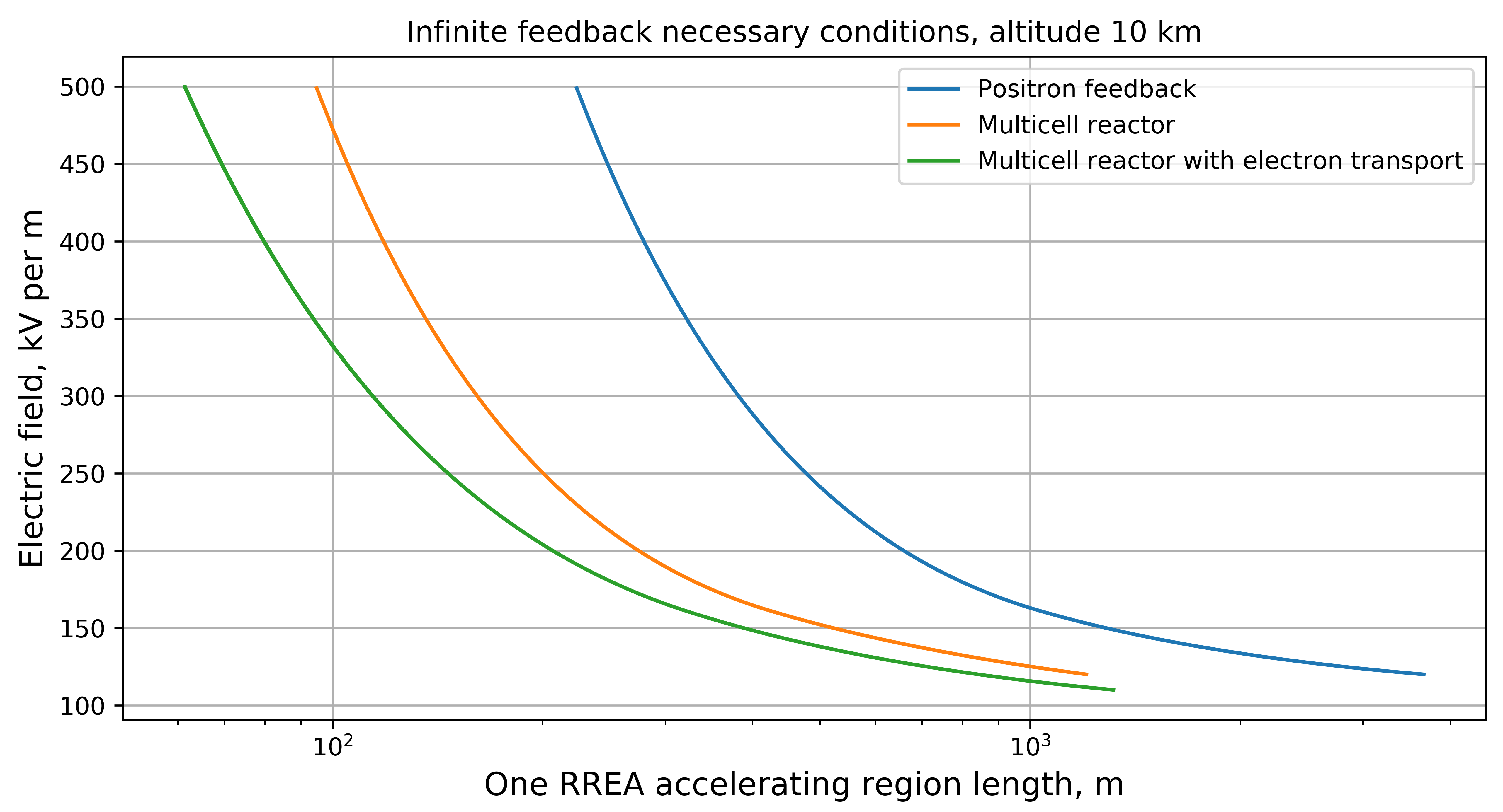}
     \caption{
    The diagram of the rate of multiplication of gamma-ray photons within the multicell reactor model for 10~km altitude. The interaction of cells is ensured by gamma-ray propagation between cells (orange curve, Formula~\ref{formula_nu_gamma}) and by the propagation of gamma-ray photons and runaway electrons for the mean distance between cells 200 m (green curve, Formula~\ref{formula_nu_rreas}, $\delta = 0.05$ due to runaway electron attenuation between cells, Appendix C). On the curves of the reactor model the local multiplication factor $\nu = 1$, which is the necessary condition for infinite reactor feedback. Above the curve, $\nu > 1$ and gamma-rays can multiply infinitely, under the curve $\nu < 1$ and the number of relativistic particles within the thunderstorm without external sources fades. The conditions required for infinite feedback in the multicell reactor model are presented in comparison with conditions required for infinite positron feedback in the Relativistic Feedback Discharge Model. Conditions for infinite positron feedback were taken from \protect\cite{Dwyer2007} with extrapolation of the feedback factor proportional to $exp(\frac{L}{\lambda_{RREA}})$, where $L$ --- critical electric field region length, $\lambda_{RREA}$ --- relativistic runaway electron avalanche e-folding length, Formula~\protect\ref{avalanche_length}.
}
    \label{local_explosion_criteria}
\end{figure}

Figure~\ref{local_explosion_criteria} presents the criterion of infinite reactor feedback in a cloud calculated for the altitude 10~km (critical electric field is approximately 90~kV/m). This criterion is derived from the necessary condition for the local multiplication factor: $\nu > 1$ (sufficient condition, when global multiplication factor is more than zero (Formula~\ref{explosion_criterion}), varies depending on reactor thunderstorm geometry, Figure~\protect\ref{models_comparison}).
It could be seen that the condition of the generation of self-sustaining gamma flash is more achievable for the case with the transport of runaway electrons between cells. Besides, the conditions required for infinite feedback in the multicell reactor model are more achievable within thunderstorms than conditions required for infinite positron feedback in Relativistic Feedback Discharge Model \protect\cite{Dwyer2007}. Figure \protect\ref{local_explosion_criteria} shows that relativistic feedback can be neglected in the multicell reactor model, as the reactor feedback dominates. This is because the relativistic feedback is limited by the rarity of bremsstrahlung gamma reversal, by the small value of electron-positron pair production cross-section (characteristic gamma-ray path before positron production for 10~km altitude is approximately 14~km \cite{NIST}), and by electron and positron reversal \protect\cite{Dwyer_2012,Stadnichuk2019,refId0}. While the reactor feedback is limited by gamma-ray transport between cells and electron reversal.

Formula \ref{formula_nu_rreas} also describes another interesting effect. If $0.5\delta e^{\frac{L}{\lambda_{RREA}}} > 1$ then the sum of a geometric progression diverges. That means that electron transport loops: on average a RREA in one cell produces a new RREA in another cell as a result of propagation of a runaway electron from one cell to another. Therefore, effectively electrons go in circles forming an infinite avalanche through the multicell reactor system. And this looped avalanche radiates gamma-rays. The condition $0.5\delta e^{\frac{L}{\lambda_{RREA}}} > 1$ describes the condition of infinite feedback in the multicell reactor due to runaway electrons transport only. It should be noted, that an important condition for the occurrence of the infinite feedback is the proximity of the cells to each other (Section \ref{Appendix_D}, Appendix C).

The proposed analysis is convenient for predicting the system behavior before the detailed simulation and provides a physical explanation for qualitative relations. Formula~\ref{explosion_criterion} might be used as a necessary condition for infinite feedback in reactor-like systems. The local multiplication factor can be estimated via Formula \ref{formula_nu_gamma} for solely gamma-ray exchange between cells and via Formula~\ref{formula_nu_rreas} for a reactor system with the exchange of gamma-rays and runaway electrons between cells. The crucial parameters of the reactor system are the electric field strength, cell length, and air density, which affect local relativistic runaway electron dynamics, influencing local gamma-ray multiplication. Air density and thunderstorm size affect macroscopic gamma-ray dynamics, its transport between cells. It should be noted that the dependence of the system behavior on the thunderstorm size is significant for the thunderstorm size less than 1.5~km, Figure~\ref{macroreactor}, while for a larger system when the size of the reactor is much bigger than the mean free path of RREA bremsstrahlung gamma-rays, the size-depending term becomes negligible, Formula~\ref{explosion_criterion}.

\section{Discussion}
\label{Discussion}

The paper analyzes the dynamics of RREAs in the thundercloud with the complex distribution of the electric field, demonstrating the impact of the new kind of positive feedback in the development of RREAs --- ``reactor feedback''. The proposed reactor model can describe both short intensive gamma-ray bursts like TGFs and long-scale particle fluxes like TGEs and gamma-ray glows, depending on the intensity of the interaction of the strong field regions in the cloud, Figure~\ref{macroreactor}.

The proposed ``cell'' concept can be considered The electrical structure of a cloud is described using the “multicell” model of field distribution containing strong field regions with uniform electric field --- “cells”. The “cell” concept provides an opportunity to study the reactor feedback in various electric structures with relatively low computational complexity. The proposed approach can be considered as the next step on the way of description of the RREA development in real thunderclouds, preceded by the model with a uniform electric field widely used in numerical modeling \cite{Dwyer2007, Skeltved2014, Chilingarian2018}.

RREA dynamics in the cylindrical electric field of a lightning leader nonuniform field are analyzed in \cite{Kutsyk2011606, Babich_2020}. The system considered in \cite{Kutsyk2011606, Babich_2020} demonstrates the feedback effect of RREA amplification influenced by the system geometry. The cylindrical structure of the electric field can be considered as the reactor structure consisting of thin radial cells with the electric field directed to the axis of the cylinder. The RREA developing in radial direction emits bremsstrahlung towards the axis and in this way amplifies RREAs in the opposite cells. The results of \cite{Kutsyk2011606,Babich_2020} supports the idea that the heterogeneity of thunderstorm electric field might lead to feedback processes in RREA dynamics, enhancing fluxes of relativistic particles in a thunderstorm. We would like to note that an arbitrary heterogeneity of the electric field can provide the reactor feedback because the radiation of the initial avalanche would easily reach other strong field regions.

The reactor model can be conveniently applied to study real clouds within the main widely used models of the cloud charge distribution. The cloud electrical structure is often described as a ``classical tripole'' or a ``dipole'', though more complicated multi-layer geometries are discussed as well \cite{Williams1989, Ette1982, Rust1996}. The widely used layered models regardless of the number of charge layers include the system of two regions with the quasi-uniform critical electric field of opposite direction. This system experiences the reactor feedback if runaway electrons accelerated in one cell move towards the cell in the opposite direction of the electric field, radiating bremsstrahlung and igniting the opposite cell \protect\cite{EGU,Dwyer2007}. Other simple geometries exhibiting the reactor feedback are ``cylindrical'' and ``spherical'' discussed in \protect\cite{Kutsyk2011606}. All mentioned geometries might lead to infinite feedback in RREA dynamics, while the electric field strength and cell length required for reactor explosion depend on the parameters of the charge structure. The modeling results shown in Figure~\ref{local_explosion_criteria} enable estimating the size of ``cells'' sufficient for infinite feedback being in range 50--500 m. The reactor model demonstrates the multiplication of avalanches if the cell is larger than the avalanche e-folding length. Investigations of the electrical structure of clouds, including direct measurements, indicate its heterogeneity. The results of the balloon- and aircraft-based measurements in thunderclouds show that the scale of heterogeneity of the electric field can lie within the estimated range of infinite feedback: 50--1000 m \cite{Marshall1995, Marshall1998_estimates, Stolzenburg2008,Tessendorf2007Multicell,Rust2005InvertedPolarity}, while the electric field on smaller scales is probably even less uniform. Numerical modeling presented in \cite{Brothers2018} shows the electrical structure consisting of a large number of charged regions (Fig.4 in \cite{Brothers2018}), which can correspond to the multicell structure with 10 cells and more with size about 0.5--1~km. Thus, we assume that the proposed mechanism of the reactor feedback can be important for the RREA development in real clouds.

The presented consideration of the multicell reactor model provides new opportunities for diagnostics of TGF and TGE mechanisms. The crucial property of the RREA development is its gamma-ray radiation pattern. The conventional RREA mechanism in the uniform electric field leads to bremsstrahlung in a narrow cone directed backward to the electric field \cite{doi:10.1029/2007JD009248}. In the multicell reactor model, the electric field in each cell might have any direction, thus the pattern can be wide-angled or even quasi-isotropic, Formula~\ref{eq:reactor_spatial}. The thundercloud with the reactor structure might radiate gamma-rays up, down, and, possibly, sideways with approximately the same brightness, depending on the electric field geometry. The analysis of the angular distribution of observed TGFs leads to the conclusion that some of TGF sources have a wider angular distribution than the directed one \cite{doi:10.1029/2008GL035906,doi:10.1029/2011JA016716,Fermi_2016}. However, a wide gamma-ray emission angle implies much more relativistic particles within thunderstorms during TGFs than with directed radiation to fit observable from space gamma-ray fluxes. In a reactor-like thunderstorm, infinite feedback is achieved via interaction between different parts of the storm, allowing the creation of a great number of relativistic particles, Figure \ref{models_comparison}.

The reactor mechanism can produce a TGF or a TGE depending on the electrical structure of the cloud, which defines the global multiplication factor, Formula \ref{eq:reactor_spatial}. The feedback effect can lead to the auto-tuning of the charge distribution, increasing the discharging for higher values of the electric field and slowing the discharging as the electrical field strength decreases.

The developed reactor model is consistent with an opportunity of lightning production by a TGF, as well as the runaway breakdown model and the relativistic feedback discharge model: reactor structure with infinite feedback leads to the significant rise of ionization and conductivity in the cloud, which can result in lightning \protect\cite{gurevich_1999,Dwyer_2003_fundamental_limit,kostinsky_2020_mechanism}. On the other hand, lightning discharge can affect the strength and direction of the local electric field, providing two opportunities for lightning to influence the development of RREAs. (1) A nearby lightning flash usually terminates a TGE or gamma-ray glow, \cite{Chilingarian2017, Wada2019}. The reactor model provides a new possible relation between a lightning flash and a RREA. Namely, a lightning flash can decrease the electric field below the critical value in some parts of the reactor, while the field in other regions would remain sufficient for the RREAs development. In other words, some strong field regions will be destroyed and some will remain, making possible the flux continuing after a lightning discharge in the cloud. The described effect might lead to TGF gamma-ray flux decay if the global multiplication factor falls below zero after the lightning discharge (Section \ref{appendix_a}, Gamma-radiation dynamics in RREA models with positive feedback) and to multi-pulse TGF if the global multiplication factor recovers to a positive value. (2) What is more, a charge transition caused by a lightning discharge might increase the heterogeneity of the electric field in the cloud, leading to TGF or gamma-ray glow initiation via the reactor feedback. The described possibility is a mechanism of energetic flux production by a lightning discharge, different from the lightning leader model. The local increase of the electric field in the reactor model may explain the TGF-like intensification of energetic flux following a gamma-ray glow reported in \cite{Wada2019}.

We demonstrate the importance of the interaction of different regions of strong fields for RREA development. We show the effect of the interaction of cells by studying the example of a ``multicell reactor'' which consists of many regions of the strong field with different directions, Figure \protect\ref{local_explosion_criteria}. We assume that the RREAs can demonstrate the reactor-like behavior in a wide variety of heterogeneous electric field structures, as far as the only necessary condition is that the bremsstrahlung of one avalanche reaches the cell where other avalanches develop. Therefore, the investigation of the thunderstorm electric field structure is crucial for understanding the physics of the RREAs and their gamma-emission.

The goal of this paper is to show that heterogeneous thunderstorm electric field geometry can lead to feedback processes in RREA dynamics by high energy particle exchange between different RREA accelerating regions. Moreover, it was shown that with the proposed feedback mechanism electric field strength required for infinite feedback, possibly responsible for TGFs, is much smaller than the values required for infinite relativistic feedback within a single accelerating region, Figure~\protect\ref{local_explosion_criteria}. However, the development of avalanches can be influenced predominantly by relativistic or reactor feedback depending on the electrical structure of the cloud.

\section{Conclusions}

In this paper, a new feedback mechanism for the dynamics of relativistic runaway electron avalanches is proposed. The ``reactor feedback'' arises in complex electric field structures due to high energy particles exchange between different strong field regions. The analysis of the multicell reactor model shows that the feedback can cause the self-sustaining development of relativistic electron avalanches, which can lead to an energetic particle flux of long duration, similar to a gamma-ray glow or TGE. Moreover, the presented mechanism with infinite feedback can produce a TGF. Based on the analytical consideration and modeling results we show that strong field regions of size 50--1000~m with different field directions are required for the reactor feedback. The distinguishing observable feature of the reactor mechanism is a wide-angle direction diagram of the resulting gamma-radiation. We demonstrate the importance of the interaction of different regions of strong fields for RREA development. We assume that the RREAs can demonstrate the reactor-like behavior in a wide variety of heterogeneous electric field structures, as far as the only necessary condition is that the bremsstrahlung of one avalanche reaches the cell where other avalanches develop. Therefore, the investigation of the thunderstorm electric field structure is crucial for understanding the physics of the RREAs and their gamma-emission.

The further study of the reactor feedback mechanism will be focused on reactor feedback behavior in organized electric field structures, for example, within a thunderstorm with two cells accelerating runaway electrons towards each other; with multiple cells oriented in a way predicted by thunderstorm electrification models and observations, etc. It requires:

\begin{itemize}
    \item More detailed microscopic modeling for a better understanding of single-cell radiation.
    \item Development of the macroscopic model capable of studying the interaction of cells for any geometry of a thunderstorm electric field.
    \item Development of new theoretical techniques to study the reactor feedback mechanism.
\end{itemize}

\acknowledgments
The work of E. Stadnichuk, A. Nozik and M. Zelenyy was funded by MIPT 5-100 academic support program. The work of E. Svechnikova was supported by a grant from the Government of the Russian Federation
(contract no. 075-15-2019-1892).
The authors thank Alexey Pozanenko for fruitful discussions of the model proposed in this article. The authors thank Makar Leonenko for the provided illustrations.
The source code and distributions of the macroscopic model implemented in Kotlin are available in the repository: https://doi.org/10.5281/zenodo.5236348


%
%

\bibliography{multicell}

\providecommand{\noopsort}[1]{}\providecommand{\singleletter}[1]{#1}%
\begin{thebibliography}{}

\bibitem [\protect \citeauthoryear {%
Agostinelli%
\ \protect \BOthers {.}}{%
Agostinelli%
\ \protect \BOthers {.}}{%
{\protect \APACyear {2003}}%
}]{%
geant4}
\APACinsertmetastar {%
geant4}%
\begin{APACrefauthors}%
Agostinelli, S.%
\BCBT {}\ \BOthersPeriod {.}
\end{APACrefauthors}%
\unskip\
\newblock
\APACrefYearMonthDay{2003}{}{}.
\newblock
{\BBOQ}\APACrefatitle {Geant4—a simulation toolkit} {Geant4—a simulation
  toolkit}.{\BBCQ}
\newblock
\APACjournalVolNumPages{Nuclear Instruments and Methods in Physics Research
  Section A: Accelerators, Spectrometers, Detectors and Associated
  Equipment}{506}{3}{250 - 303}.
\newblock
\begin{APACrefURL}
  \url{http://www.sciencedirect.com/science/article/pii/S0168900203013688}
  \end{APACrefURL}
\newblock
\begin{APACrefDOI} \doi{https://doi.org/10.1016/S0168-9002(03)01368-8}
  \end{APACrefDOI}
\PrintBackRefs{\CurrentBib}

\bibitem [\protect \citeauthoryear {%
Babich%
}{%
Babich%
}{%
{\protect \APACyear {2020}}%
}]{%
Babich_2020}
\APACinsertmetastar {%
Babich_2020}%
\begin{APACrefauthors}%
Babich, L\BPBI P.%
\end{APACrefauthors}%
\unskip\
\newblock
\APACrefYearMonthDay{2020}{dec}{}.
\newblock
{\BBOQ}\APACrefatitle {Relativistic runaway electron avalanche} {Relativistic
  runaway electron avalanche}.{\BBCQ}
\newblock
\APACjournalVolNumPages{Physics-Uspekhi}{63}{12}{1188--1218}.
\newblock
\begin{APACrefURL} \url{https://doi.org/10.3367/ufne.2020.04.038747}
  \end{APACrefURL}
\newblock
\begin{APACrefDOI} \doi{10.3367/ufne.2020.04.038747} \end{APACrefDOI}
\PrintBackRefs{\CurrentBib}

\bibitem [\protect \citeauthoryear {%
Berger%
\ \protect \BOthers {.}}{%
Berger%
\ \protect \BOthers {.}}{%
{\protect \APACyear {2010}}%
}]{%
NIST}
\APACinsertmetastar {%
NIST}%
\begin{APACrefauthors}%
Berger, M.%
, Hubbell, J.%
, Seltzer, S.%
, Chang, J.%
, Coursey, J.%
, Sukumar, R.%
\BDBL {}Olsen, K.%
\end{APACrefauthors}%
\unskip\
\newblock
\APACrefYearMonthDay{2010}{nov}{}.
\newblock
{\BBOQ}\APACrefatitle {XCOM: Photon Cross Sections Database} {Xcom: Photon
  cross sections database}.{\BBCQ}
\newblock
\APACjournalVolNumPages{Physics-Uspekhi}{}{}{}.
\newblock
\begin{APACrefURL} \url{https://dx.doi.org/10.18434/T48G6X} \end{APACrefURL}
\newblock
\begin{APACrefDOI} \doi{10.18434/T48G6X} \end{APACrefDOI}
\PrintBackRefs{\CurrentBib}

\bibitem [\protect \citeauthoryear {%
Brothers%
, Bruning%
\BCBL {}\ \BBA {} Mansell%
}{%
Brothers%
\ \protect \BOthers {.}}{%
{\protect \APACyear {2018}}%
}]{%
Brothers2018}
\APACinsertmetastar {%
Brothers2018}%
\begin{APACrefauthors}%
Brothers, M.%
, Bruning, E.%
\BCBL {}\ \BBA {} Mansell, E.%
\end{APACrefauthors}%
\unskip\
\newblock
\APACrefYearMonthDay{2018}{07}{}.
\newblock
{\BBOQ}\APACrefatitle {Investigating the Relative Contributions of Charge
  Deposition and Turbulence in Organizing Charge within a Thunderstorm}
  {Investigating the relative contributions of charge deposition and turbulence
  in organizing charge within a thunderstorm}.{\BBCQ}
\newblock
\APACjournalVolNumPages{Journal of the Atmospheric Sciences}{75}{}{}.
\newblock
\begin{APACrefDOI} \doi{10.1175/JAS-D-18-0007.1} \end{APACrefDOI}
\PrintBackRefs{\CurrentBib}

\bibitem [\protect \citeauthoryear {%
Chilingarian%
}{%
Chilingarian%
}{%
{\protect \APACyear {2011}}%
}]{%
Chilingarian_2011_natural_accelerator}
\APACinsertmetastar {%
Chilingarian_2011_natural_accelerator}%
\begin{APACrefauthors}%
Chilingarian, A.%
\end{APACrefauthors}%
\unskip\
\newblock
\APACrefYearMonthDay{2011}{03}{}.
\newblock
{\BBOQ}\APACrefatitle {Particle bursts from thunderclouds: Natural particle
  accelerators above our heads} {Particle bursts from thunderclouds: Natural
  particle accelerators above our heads}.{\BBCQ}
\newblock
\APACjournalVolNumPages{Phys. Rev. D}{83}{}{}.
\newblock
\begin{APACrefDOI} \doi{10.1103/PhysRevD.83.062001} \end{APACrefDOI}
\PrintBackRefs{\CurrentBib}

\bibitem [\protect \citeauthoryear {%
Chilingarian%
, Hovsepyan%
, Soghomonyan%
, Zazyan%
\BCBL {}\ \BBA {} Zelenyy%
}{%
Chilingarian%
\ \protect \BOthers {.}}{%
{\protect \APACyear {2018}}%
}]{%
Chilingarian2018}
\APACinsertmetastar {%
Chilingarian2018}%
\begin{APACrefauthors}%
Chilingarian, A.%
, Hovsepyan, G.%
, Soghomonyan, S.%
, Zazyan, M.%
\BCBL {}\ \BBA {} Zelenyy, M.%
\end{APACrefauthors}%
\unskip\
\newblock
\APACrefYearMonthDay{2018}{10}{}.
\newblock
{\BBOQ}\APACrefatitle {Structures of the intracloud electric field supporting
  origin of long-lasting thunderstorm ground enhancements} {Structures of the
  intracloud electric field supporting origin of long-lasting thunderstorm
  ground enhancements}.{\BBCQ}
\newblock
\APACjournalVolNumPages{Physical Review D}{98}{}{}.
\newblock
\begin{APACrefDOI} \doi{10.1103/PhysRevD.98.082001} \end{APACrefDOI}
\PrintBackRefs{\CurrentBib}

\bibitem [\protect \citeauthoryear {%
Chilingarian%
\ \protect \BOthers {.}}{%
Chilingarian%
\ \protect \BOthers {.}}{%
{\protect \APACyear {2017}}%
}]{%
Chilingarian2017}
\APACinsertmetastar {%
Chilingarian2017}%
\begin{APACrefauthors}%
Chilingarian, A.%
, Khanikyants, Y.%
, Mareev, E.%
, Pokhsraryan, D.%
, Rakov, V.%
\BCBL {}\ \BBA {} Soghomonyan, S.%
\end{APACrefauthors}%
\unskip\
\newblock
\APACrefYearMonthDay{2017}{07}{}.
\newblock
{\BBOQ}\APACrefatitle {Types of lightning discharges that abruptly terminate
  enhanced fluxes of energetic radiation and particles observed at ground
  level: Types of TGE-terminating lightning flashes} {Types of lightning
  discharges that abruptly terminate enhanced fluxes of energetic radiation and
  particles observed at ground level: Types of tge-terminating lightning
  flashes}.{\BBCQ}
\newblock
\APACjournalVolNumPages{Journal of Geophysical Research: Atmospheres}{122}{}{}.
\newblock
\begin{APACrefDOI} \doi{10.1002/2017JD026744} \end{APACrefDOI}
\PrintBackRefs{\CurrentBib}

\bibitem [\protect \citeauthoryear {%
J.~Dwyer%
, Smith%
\BCBL {}\ \BBA {} Cummer%
}{%
J.~Dwyer%
\ \protect \BOthers {.}}{%
{\protect \APACyear {2012}}%
}]{%
Dwyer2012_phenomena}
\APACinsertmetastar {%
Dwyer2012_phenomena}%
\begin{APACrefauthors}%
Dwyer, J.%
, Smith, D.%
\BCBL {}\ \BBA {} Cummer, S.%
\end{APACrefauthors}%
\unskip\
\newblock
\APACrefYearMonthDay{2012}{}{}.
\newblock
{\BBOQ}\APACrefatitle {High-Energy Atmospheric Physics: Terrestrial Gamma-Ray
  Flashes and Related Phenomena} {High-energy atmospheric physics: Terrestrial
  gamma-ray flashes and related phenomena}.{\BBCQ}
\newblock
\APACjournalVolNumPages{Space Sci. Rev.}{177}{}{133--196}.
\PrintBackRefs{\CurrentBib}

\bibitem [\protect \citeauthoryear {%
J\BPBI R.~Dwyer%
}{%
J\BPBI R.~Dwyer%
}{%
{\protect \APACyear {2003}}%
}]{%
Dwyer_2003_fundamental_limit}
\APACinsertmetastar {%
Dwyer_2003_fundamental_limit}%
\begin{APACrefauthors}%
Dwyer, J\BPBI R.%
\end{APACrefauthors}%
\unskip\
\newblock
\APACrefYearMonthDay{2003}{}{}.
\newblock
{\BBOQ}\APACrefatitle {A fundamental limit on electric fields in air} {A
  fundamental limit on electric fields in air}.{\BBCQ}
\newblock
\APACjournalVolNumPages{Geophysical Research Letters}{30}{20}{}.
\newblock
\begin{APACrefURL}
  \url{https://agupubs.onlinelibrary.wiley.com/doi/abs/10.1029/2003GL017781}
  \end{APACrefURL}
\newblock
\begin{APACrefDOI} \doi{10.1029/2003GL017781} \end{APACrefDOI}
\PrintBackRefs{\CurrentBib}

\bibitem [\protect \citeauthoryear {%
J\BPBI R.~Dwyer%
}{%
J\BPBI R.~Dwyer%
}{%
{\protect \APACyear {2007}}%
}]{%
Dwyer2007}
\APACinsertmetastar {%
Dwyer2007}%
\begin{APACrefauthors}%
Dwyer, J\BPBI R.%
\end{APACrefauthors}%
\unskip\
\newblock
\APACrefYearMonthDay{2007}{}{}.
\newblock
{\BBOQ}\APACrefatitle {Relativistic breakdown in planetary atmospheres}
  {Relativistic breakdown in planetary atmospheres}.{\BBCQ}
\newblock
\APACjournalVolNumPages{Physics of Plasmas}{14}{4}{042901}.
\newblock
\begin{APACrefURL} \url{https://doi.org/10.1063/1.2709652} \end{APACrefURL}
\newblock
\begin{APACrefDOI} \doi{10.1063/1.2709652} \end{APACrefDOI}
\PrintBackRefs{\CurrentBib}

\bibitem [\protect \citeauthoryear {%
J\BPBI R.~Dwyer%
}{%
J\BPBI R.~Dwyer%
}{%
{\protect \APACyear {2008}}%
}]{%
doi:10.1029/2007JD009248}
\APACinsertmetastar {%
doi:10.1029/2007JD009248}%
\begin{APACrefauthors}%
Dwyer, J\BPBI R.%
\end{APACrefauthors}%
\unskip\
\newblock
\APACrefYearMonthDay{2008}{}{}.
\newblock
{\BBOQ}\APACrefatitle {Source mechanisms of terrestrial gamma-ray flashes}
  {Source mechanisms of terrestrial gamma-ray flashes}.{\BBCQ}
\newblock
\APACjournalVolNumPages{Journal of Geophysical Research:
  Atmospheres}{113}{D10}{}.
\newblock
\begin{APACrefDOI} \doi{10.1029/2007JD009248} \end{APACrefDOI}
\PrintBackRefs{\CurrentBib}

\bibitem [\protect \citeauthoryear {%
J\BPBI R.~Dwyer%
}{%
J\BPBI R.~Dwyer%
}{%
{\protect \APACyear {2012}}%
}]{%
Dwyer_2012}
\APACinsertmetastar {%
Dwyer_2012}%
\begin{APACrefauthors}%
Dwyer, J\BPBI R.%
\end{APACrefauthors}%
\unskip\
\newblock
\APACrefYearMonthDay{2012}{}{}.
\newblock
{\BBOQ}\APACrefatitle {The relativistic feedback discharge model of terrestrial
  gamma ray flashes} {The relativistic feedback discharge model of terrestrial
  gamma ray flashes}.{\BBCQ}
\newblock
\APACjournalVolNumPages{Journal of Geophysical Research: Space
  Physics}{117}{A2}{}.
\newblock
\begin{APACrefURL}
  \url{https://agupubs.onlinelibrary.wiley.com/doi/abs/10.1029/2011JA017160}
  \end{APACrefURL}
\newblock
\begin{APACrefDOI} \doi{https://doi.org/10.1029/2011JA017160} \end{APACrefDOI}
\PrintBackRefs{\CurrentBib}

\bibitem [\protect \citeauthoryear {%
J\BPBI R.~Dwyer%
\ \BBA {} Cummer%
}{%
J\BPBI R.~Dwyer%
\ \BBA {} Cummer%
}{%
{\protect \APACyear {2013}}%
}]{%
doi:10.1002/jgra.50188}
\APACinsertmetastar {%
doi:10.1002/jgra.50188}%
\begin{APACrefauthors}%
Dwyer, J\BPBI R.%
\BCBT {}\ \BBA {} Cummer, S\BPBI A.%
\end{APACrefauthors}%
\unskip\
\newblock
\APACrefYearMonthDay{2013}{}{}.
\newblock
{\BBOQ}\APACrefatitle {Radio emissions from terrestrial gamma-ray flashes}
  {Radio emissions from terrestrial gamma-ray flashes}.{\BBCQ}
\newblock
\APACjournalVolNumPages{Journal of Geophysical Research: Space Physics}{}{}{}.
\PrintBackRefs{\CurrentBib}

\bibitem [\protect \citeauthoryear {%
Ette%
\ \BBA {} Olaofe%
}{%
Ette%
\ \BBA {} Olaofe%
}{%
{\protect \APACyear {1982}}%
}]{%
Ette1982}
\APACinsertmetastar {%
Ette1982}%
\begin{APACrefauthors}%
Ette, A.%
\BCBT {}\ \BBA {} Olaofe, G.%
\end{APACrefauthors}%
\unskip\
\newblock
\APACrefYearMonthDay{1982}{01}{}.
\newblock
{\BBOQ}\APACrefatitle {Theoretical field configurations for thundercloud models
  with volume charge distributions} {Theoretical field configurations for
  thundercloud models with volume charge distributions}.{\BBCQ}
\newblock
\APACjournalVolNumPages{Pure and Applied Geophysics}{120}{}{117-122}.
\newblock
\begin{APACrefDOI} \doi{10.1007/BF00879431} \end{APACrefDOI}
\PrintBackRefs{\CurrentBib}

\bibitem [\protect \citeauthoryear {%
Fishman%
\ \protect \BOthers {.}}{%
Fishman%
\ \protect \BOthers {.}}{%
{\protect \APACyear {1994}}%
}]{%
Fishman1994}
\APACinsertmetastar {%
Fishman1994}%
\begin{APACrefauthors}%
Fishman, G.%
, Bhat, P.%
, Mallozzi, R.%
, Horack, L.%
, Koshut, T.%
, Kouveliotou, C.%
\BDBL {}Christian, H.%
\end{APACrefauthors}%
\unskip\
\newblock
\APACrefYearMonthDay{1994}{}{}.
\newblock
{\BBOQ}\APACrefatitle {Discovery of Intense Gamma-Ray Flashes of Atmospheric
  Origin} {Discovery of intense gamma-ray flashes of atmospheric
  origin}.{\BBCQ}
\newblock
\APACjournalVolNumPages{Science}{264}{}{1313--1316}.
\PrintBackRefs{\CurrentBib}

\bibitem [\protect \citeauthoryear {%
Gjesteland%
\ \protect \BOthers {.}}{%
Gjesteland%
\ \protect \BOthers {.}}{%
{\protect \APACyear {2011}}%
}]{%
doi:10.1029/2011JA016716}
\APACinsertmetastar {%
doi:10.1029/2011JA016716}%
\begin{APACrefauthors}%
Gjesteland, T.%
, Østgaard, N.%
, Collier, A\BPBI B.%
, Carlson, B\BPBI E.%
, Cohen, M\BPBI B.%
\BCBL {}\ \BBA {} Lehtinen, N\BPBI G.%
\end{APACrefauthors}%
\unskip\
\newblock
\APACrefYearMonthDay{2011}{}{}.
\newblock
{\BBOQ}\APACrefatitle {Confining the angular distribution of terrestrial gamma
  ray flash emission} {Confining the angular distribution of terrestrial gamma
  ray flash emission}.{\BBCQ}
\newblock
\APACjournalVolNumPages{Journal of Geophysical Research: Space
  Physics}{116}{A11}{}.
\newblock
\begin{APACrefDOI} \doi{10.1029/2011JA016716} \end{APACrefDOI}
\PrintBackRefs{\CurrentBib}

\bibitem [\protect \citeauthoryear {%
A.~Gurevich%
\ \protect \BOthers {.}}{%
A.~Gurevich%
\ \protect \BOthers {.}}{%
{\protect \APACyear {2016}}%
}]{%
Gurevich2016_tien_shan}
\APACinsertmetastar {%
Gurevich2016_tien_shan}%
\begin{APACrefauthors}%
Gurevich, A.%
, Almenova, A.%
, Antonova, V.%
, Chubenko, A.%
, Karashtin, A.%
, Kryakunova, O.%
\BDBL {}Zybin, K.%
\end{APACrefauthors}%
\unskip\
\newblock
\APACrefYearMonthDay{2016}{07}{}.
\newblock
{\BBOQ}\APACrefatitle {Observations of high-energy radiation during
  thunderstorms at {Tien-Shan}} {Observations of high-energy radiation during
  thunderstorms at {Tien-Shan}}.{\BBCQ}
\newblock
\APACjournalVolNumPages{Physical Review D}{94}{}{}.
\newblock
\begin{APACrefDOI} \doi{10.1103/PhysRevD.94.023003} \end{APACrefDOI}
\PrintBackRefs{\CurrentBib}

\bibitem [\protect \citeauthoryear {%
A.~Gurevich%
, Milikh%
\BCBL {}\ \BBA {} Roussel-Dupré%
}{%
A.~Gurevich%
\ \protect \BOthers {.}}{%
{\protect \APACyear {1992}}%
}]{%
Gurevich1992}
\APACinsertmetastar {%
Gurevich1992}%
\begin{APACrefauthors}%
Gurevich, A.%
, Milikh, G.%
\BCBL {}\ \BBA {} Roussel-Dupré, R.%
\end{APACrefauthors}%
\unskip\
\newblock
\APACrefYearMonthDay{1992}{}{}.
\newblock
{\BBOQ}\APACrefatitle {Recovering of the energy spectra of electrons and gamma
  rays coming from the thunderclouds} {Recovering of the energy spectra of
  electrons and gamma rays coming from the thunderclouds}.{\BBCQ}
\newblock
\APACjournalVolNumPages{Phys. Lett. A}{165}{}{463--468}.
\PrintBackRefs{\CurrentBib}

\bibitem [\protect \citeauthoryear {%
A.~Gurevich%
, Zybin%
\BCBL {}\ \BBA {} Roussel-Dupre%
}{%
A.~Gurevich%
\ \protect \BOthers {.}}{%
{\protect \APACyear {1999}}%
}]{%
gurevich_1999}
\APACinsertmetastar {%
gurevich_1999}%
\begin{APACrefauthors}%
Gurevich, A.%
, Zybin, K.%
\BCBL {}\ \BBA {} Roussel-Dupre, R.%
\end{APACrefauthors}%
\unskip\
\newblock
\APACrefYearMonthDay{1999}{04}{}.
\newblock
{\BBOQ}\APACrefatitle {Lightning initiation by simultaneous effect of runaway
  breakdown and cosmic ray showers} {Lightning initiation by simultaneous
  effect of runaway breakdown and cosmic ray showers}.{\BBCQ}
\newblock
\APACjournalVolNumPages{Physics Letters A}{254}{}{79-87}.
\newblock
\begin{APACrefDOI} \doi{10.1016/S0375-9601(99)00091-2} \end{APACrefDOI}
\PrintBackRefs{\CurrentBib}

\bibitem [\protect \citeauthoryear {%
A\BPBI V.~Gurevich%
\ \BBA {} Zybin%
}{%
A\BPBI V.~Gurevich%
\ \BBA {} Zybin%
}{%
{\protect \APACyear {2001}}%
}]{%
Gurevich_2001}
\APACinsertmetastar {%
Gurevich_2001}%
\begin{APACrefauthors}%
Gurevich, A\BPBI V.%
\BCBT {}\ \BBA {} Zybin, K\BPBI P.%
\end{APACrefauthors}%
\unskip\
\newblock
\APACrefYearMonthDay{2001}{}{}.
\newblock
{\BBOQ}\APACrefatitle {Runaway breakdown and electric discharges in
  thunderstorms} {Runaway breakdown and electric discharges in
  thunderstorms}.{\BBCQ}
\newblock
\APACjournalVolNumPages{Uspekhi Fizicheskikh Nauk ({UFN})
  Journal}{44}{11}{1119--1140}.
\newblock
\begin{APACrefDOI} \doi{10.1070/pu2001v044n11abeh000939} \end{APACrefDOI}
\PrintBackRefs{\CurrentBib}

\bibitem [\protect \citeauthoryear {%
Hazelton%
\ \protect \BOthers {.}}{%
Hazelton%
\ \protect \BOthers {.}}{%
{\protect \APACyear {2009}}%
}]{%
doi:10.1029/2008GL035906}
\APACinsertmetastar {%
doi:10.1029/2008GL035906}%
\begin{APACrefauthors}%
Hazelton, B\BPBI J.%
, Grefenstette, B\BPBI W.%
, Smith, D\BPBI M.%
, Dwyer, J\BPBI R.%
, Shao, X\BHBI M.%
, Cummer, S\BPBI A.%
\BDBL {}Holzworth, R\BPBI H.%
\end{APACrefauthors}%
\unskip\
\newblock
\APACrefYearMonthDay{2009}{}{}.
\newblock
{\BBOQ}\APACrefatitle {Spectral dependence of terrestrial gamma-ray flashes on
  source distance} {Spectral dependence of terrestrial gamma-ray flashes on
  source distance}.{\BBCQ}
\newblock
\APACjournalVolNumPages{Geophysical Research Letters}{36}{1}{}.
\newblock
\begin{APACrefDOI} \doi{10.1029/2008GL035906} \end{APACrefDOI}
\PrintBackRefs{\CurrentBib}

\bibitem [\protect \citeauthoryear {%
Khamitov%
, Nozik%
, Stadnichuk%
, Svechnikova%
\BCBL {}\ \BBA {} Zelenyi%
}{%
Khamitov%
\ \protect \BOthers {.}}{%
{\protect \APACyear {2020}}%
}]{%
Khamiton2020}
\APACinsertmetastar {%
Khamiton2020}%
\begin{APACrefauthors}%
Khamitov, T.%
, Nozik, A.%
, Stadnichuk, E.%
, Svechnikova, E.%
\BCBL {}\ \BBA {} Zelenyi, M.%
\end{APACrefauthors}%
\unskip\
\newblock
\APACrefYearMonthDay{2020}{nov}{}.
\newblock
{\BBOQ}\APACrefatitle {Estimation of number of runaway electrons per avalanche
  in Earth{\textquotesingle}s atmosphere} {Estimation of number of runaway
  electrons per avalanche in earth{\textquotesingle}s atmosphere}.{\BBCQ}
\newblock
\APACjournalVolNumPages{{EPL} (Europhysics Letters)}{132}{3}{35001}.
\newblock
\begin{APACrefURL} \url{https://doi.org/10.1209/0295-5075/132/35001}
  \end{APACrefURL}
\newblock
\begin{APACrefDOI} \doi{10.1209/0295-5075/132/35001} \end{APACrefDOI}
\PrintBackRefs{\CurrentBib}

\bibitem [\protect \citeauthoryear {%
Kostinskiy%
, Marshall%
\BCBL {}\ \BBA {} Stolzenburg%
}{%
Kostinskiy%
\ \protect \BOthers {.}}{%
{\protect \APACyear {2020}}%
}]{%
kostinsky_2020_mechanism}
\APACinsertmetastar {%
kostinsky_2020_mechanism}%
\begin{APACrefauthors}%
Kostinskiy, A.%
, Marshall, T.%
\BCBL {}\ \BBA {} Stolzenburg, M.%
\end{APACrefauthors}%
\unskip\
\newblock
\APACrefYearMonthDay{2020}{11}{}.
\newblock
{\BBOQ}\APACrefatitle {The Mechanism of the Origin and Development of Lightning
  From Initiating Event to Initial Breakdown Pulses} {The mechanism of the
  origin and development of lightning from initiating event to initial
  breakdown pulses}.{\BBCQ}
\newblock
\APACjournalVolNumPages{Journal of Geophysical Research Atmospheres}{125}{}{}.
\newblock
\begin{APACrefDOI} \doi{10.1029/2020JD033191} \end{APACrefDOI}
\PrintBackRefs{\CurrentBib}

\bibitem [\protect \citeauthoryear {%
Kutsyk%
, Babich%
\BCBL {}\ \BBA {} Donskoi%
}{%
Kutsyk%
\ \protect \BOthers {.}}{%
{\protect \APACyear {2011}}%
}]{%
Kutsyk2011606}
\APACinsertmetastar {%
Kutsyk2011606}%
\begin{APACrefauthors}%
Kutsyk, I.%
, Babich, L.%
\BCBL {}\ \BBA {} Donskoi, E.%
\end{APACrefauthors}%
\unskip\
\newblock
\APACrefYearMonthDay{2011}{}{}.
\newblock
{\BBOQ}\APACrefatitle {Self-sustained relativistic-runaway-electron avalanches
  in the transverse field of lightning leader as sources of terrestrial
  gamma-ray flashes} {Self-sustained relativistic-runaway-electron avalanches
  in the transverse field of lightning leader as sources of terrestrial
  gamma-ray flashes}.{\BBCQ}
\newblock
\APACjournalVolNumPages{JETP Letters}{94}{8}{606-609}.
\newblock
\begin{APACrefURL}
  \url{https://www.scopus.com/inward/record.uri?eid=2-s2.0-84155181583&doi=10.1134%2fS0021364011200094&partnerID=40&md5=8d0c63b0d900fbef4d635cc55a9851f0}
  \end{APACrefURL}
\newblock
\APACrefnote{cited By 10}
\newblock
\begin{APACrefDOI} \doi{10.1134/S0021364011200094} \end{APACrefDOI}
\PrintBackRefs{\CurrentBib}

\bibitem [\protect \citeauthoryear {%
Lehtinen%
, Bell%
\BCBL {}\ \BBA {} Inan%
}{%
Lehtinen%
\ \protect \BOthers {.}}{%
{\protect \APACyear {1999}}%
}]{%
Lehtinen1999}
\APACinsertmetastar {%
Lehtinen1999}%
\begin{APACrefauthors}%
Lehtinen, N\BPBI G.%
, Bell, T\BPBI F.%
\BCBL {}\ \BBA {} Inan, U\BPBI S.%
\end{APACrefauthors}%
\unskip\
\newblock
\APACrefYearMonthDay{1999}{}{}.
\newblock
{\BBOQ}\APACrefatitle {Monte Carlo simulation of runaway MeV electron breakdown
  with application to red sprites and terrestrial gamma ray flashes} {Monte
  carlo simulation of runaway mev electron breakdown with application to red
  sprites and terrestrial gamma ray flashes}.{\BBCQ}
\newblock
\APACjournalVolNumPages{Journal of Geophysical Research: Space
  Physics}{104}{A11}{24699-24712}.
\newblock
\begin{APACrefURL}
  \url{https://agupubs.onlinelibrary.wiley.com/doi/abs/10.1029/1999JA900335}
  \end{APACrefURL}
\newblock
\begin{APACrefDOI} \doi{https://doi.org/10.1029/1999JA900335} \end{APACrefDOI}
\PrintBackRefs{\CurrentBib}

\bibitem [\protect \citeauthoryear {%
Mailyan%
\ \protect \BOthers {.}}{%
Mailyan%
\ \protect \BOthers {.}}{%
{\protect \APACyear {2016}}%
}]{%
Fermi_2016}
\APACinsertmetastar {%
Fermi_2016}%
\begin{APACrefauthors}%
Mailyan, B\BPBI G.%
, Briggs, M\BPBI S.%
, Cramer, E\BPBI S.%
, Fitzpatrick, G.%
, Roberts, O\BPBI J.%
, Stanbro, M.%
\BDBL {}Dwyer, J\BPBI R.%
\end{APACrefauthors}%
\unskip\
\newblock
\APACrefYearMonthDay{2016}{}{}.
\newblock
{\BBOQ}\APACrefatitle {The spectroscopy of individual terrestrial gamma-ray
  flashes: Constraining the source properties} {The spectroscopy of individual
  terrestrial gamma-ray flashes: Constraining the source properties}.{\BBCQ}
\newblock
\APACjournalVolNumPages{Journal of Geophysical Research: Space
  Physics}{121}{11}{11,346-11,363}.
\newblock
\begin{APACrefURL}
  \url{https://agupubs.onlinelibrary.wiley.com/doi/abs/10.1002/2016JA022702}
  \end{APACrefURL}
\newblock
\begin{APACrefDOI} \doi{https://doi.org/10.1002/2016JA022702} \end{APACrefDOI}
\PrintBackRefs{\CurrentBib}

\bibitem [\protect \citeauthoryear {%
Marshall%
\ \protect \BOthers {.}}{%
Marshall%
\ \protect \BOthers {.}}{%
{\protect \APACyear {1995}}%
}]{%
Marshall1995}
\APACinsertmetastar {%
Marshall1995}%
\begin{APACrefauthors}%
Marshall, T.%
, Rison, W.%
, Rust, W.%
, Stolzenburg, M.%
, Willett, J.%
\BCBL {}\ \BBA {} Winn, W.%
\end{APACrefauthors}%
\unskip\
\newblock
\APACrefYearMonthDay{1995}{10}{}.
\newblock
{\BBOQ}\APACrefatitle {Rocket and balloon observations of electric field in two
  thunderstorms} {Rocket and balloon observations of electric field in two
  thunderstorms}.{\BBCQ}
\newblock
\APACjournalVolNumPages{Journal of Geophysical Research}{100}{}{20815-20828}.
\newblock
\begin{APACrefDOI} \doi{10.1029/95JD01877} \end{APACrefDOI}
\PrintBackRefs{\CurrentBib}

\bibitem [\protect \citeauthoryear {%
Marshall%
\ \BBA {} Stolzenburg%
}{%
Marshall%
\ \BBA {} Stolzenburg%
}{%
{\protect \APACyear {1998}}%
}]{%
Marshall1998_estimates}
\APACinsertmetastar {%
Marshall1998_estimates}%
\begin{APACrefauthors}%
Marshall, T.%
\BCBT {}\ \BBA {} Stolzenburg, M.%
\end{APACrefauthors}%
\unskip\
\newblock
\APACrefYearMonthDay{1998}{08}{}.
\newblock
{\BBOQ}\APACrefatitle {Estimates of cloud charge densities in thunderstorms}
  {Estimates of cloud charge densities in thunderstorms}.{\BBCQ}
\newblock
\APACjournalVolNumPages{Journal of Geophysical Research}{1031}{}{19769-19776}.
\newblock
\begin{APACrefDOI} \doi{10.1029/98JD01674} \end{APACrefDOI}
\PrintBackRefs{\CurrentBib}

\bibitem [\protect \citeauthoryear {%
Moss%
, Pasko%
, Liu%
\BCBL {}\ \BBA {} Veronis%
}{%
Moss%
\ \protect \BOthers {.}}{%
{\protect \APACyear {2006}}%
}]{%
https://doi.org/10.1029/2005JA011350}
\APACinsertmetastar {%
https://doi.org/10.1029/2005JA011350}%
\begin{APACrefauthors}%
Moss, G\BPBI D.%
, Pasko, V\BPBI P.%
, Liu, N.%
\BCBL {}\ \BBA {} Veronis, G.%
\end{APACrefauthors}%
\unskip\
\newblock
\APACrefYearMonthDay{2006}{}{}.
\newblock
{\BBOQ}\APACrefatitle {{Monte Carlo} model for analysis of thermal runaway
  electrons in streamer tips in transient luminous events and streamer zones of
  lightning leaders} {{Monte Carlo} model for analysis of thermal runaway
  electrons in streamer tips in transient luminous events and streamer zones of
  lightning leaders}.{\BBCQ}
\newblock
\APACjournalVolNumPages{Journal of Geophysical Research: Space
  Physics}{111}{A2}{}.
\newblock
\begin{APACrefURL}
  \url{https://agupubs.onlinelibrary.wiley.com/doi/abs/10.1029/2005JA011350}
  \end{APACrefURL}
\newblock
\begin{APACrefDOI} \doi{https://doi.org/10.1029/2005JA011350} \end{APACrefDOI}
\PrintBackRefs{\CurrentBib}

\bibitem [\protect \citeauthoryear {%
Nozik%
}{%
Nozik%
}{%
{\protect \APACyear {2019}}%
}]{%
Kotlin}
\APACinsertmetastar {%
Kotlin}%
\begin{APACrefauthors}%
Nozik, A.%
\end{APACrefauthors}%
\unskip\
\newblock
\APACrefYearMonthDay{2019}{}{}.
\newblock
{\BBOQ}\APACrefatitle {Kotlin language for science and {Kmath} library} {Kotlin
  language for science and {Kmath} library}.{\BBCQ}
\newblock
\APACjournalVolNumPages{AIP Conference Proceedings}{2163}{1}{040004}.
\newblock
\begin{APACrefURL} \url{https://aip.scitation.org/doi/abs/10.1063/1.5130103}
  \end{APACrefURL}
\newblock
\begin{APACrefDOI} \doi{10.1063/1.5130103} \end{APACrefDOI}
\PrintBackRefs{\CurrentBib}

\bibitem [\protect \citeauthoryear {%
Rust%
\ \protect \BOthers {.}}{%
Rust%
\ \protect \BOthers {.}}{%
{\protect \APACyear {2005}}%
}]{%
Rust2005InvertedPolarity}
\APACinsertmetastar {%
Rust2005InvertedPolarity}%
\begin{APACrefauthors}%
Rust, W.%
, MacGorman, D.%
, Bruning, E.%
, Weiss, S.%
, Krehbiel, P.%
, Thomas, R.%
\BDBL {}Harlin, J.%
\end{APACrefauthors}%
\unskip\
\newblock
\APACrefYearMonthDay{2005}{07}{}.
\newblock
{\BBOQ}\APACrefatitle {Inverted-polarity electrical structures in thunderstorms
  in the Severe Thunderstorm Electrification and Precipitation Study (STEPS)}
  {Inverted-polarity electrical structures in thunderstorms in the severe
  thunderstorm electrification and precipitation study (steps)}.{\BBCQ}
\newblock
\APACjournalVolNumPages{Atmospheric Research}{76}{}{247-271}.
\newblock
\begin{APACrefDOI} \doi{10.1016/j.atmosres.2004.11.029} \end{APACrefDOI}
\PrintBackRefs{\CurrentBib}

\bibitem [\protect \citeauthoryear {%
Rust%
\ \BBA {} Marshall%
}{%
Rust%
\ \BBA {} Marshall%
}{%
{\protect \APACyear {1996}}%
}]{%
Rust1996}
\APACinsertmetastar {%
Rust1996}%
\begin{APACrefauthors}%
Rust, W.%
\BCBT {}\ \BBA {} Marshall, T.%
\end{APACrefauthors}%
\unskip\
\newblock
\APACrefYearMonthDay{1996}{10}{}.
\newblock
{\BBOQ}\APACrefatitle {On abandoning the thunderstorm tripole-charge paradigm}
  {On abandoning the thunderstorm tripole-charge paradigm}.{\BBCQ}
\newblock
\APACjournalVolNumPages{Journal of Geophysical Research}{101}{}{23499-23504}.
\newblock
\begin{APACrefDOI} \doi{10.1029/96JD01802} \end{APACrefDOI}
\PrintBackRefs{\CurrentBib}

\bibitem [\protect \citeauthoryear {%
Sarria%
\ \protect \BOthers {.}}{%
Sarria%
\ \protect \BOthers {.}}{%
{\protect \APACyear {2021}}%
}]{%
asim_spectrum}
\APACinsertmetastar {%
asim_spectrum}%
\begin{APACrefauthors}%
Sarria, D.%
, Østgaard, N.%
, Kochkin, P.%
, Lehtinen, N.%
, Mezentsev, A.%
, Marisaldi, M.%
\BDBL {}Eyles, C.%
\end{APACrefauthors}%
\unskip\
\newblock
\APACrefYearMonthDay{2021}{}{}.
\newblock
{\BBOQ}\APACrefatitle {Constraining Spectral Models of a Terrestrial Gamma-Ray
  Flash From a Terrestrial Electron Beam Observation by the Atmosphere-Space
  Interactions Monitor} {Constraining spectral models of a terrestrial
  gamma-ray flash from a terrestrial electron beam observation by the
  atmosphere-space interactions monitor}.{\BBCQ}
\newblock
\APACjournalVolNumPages{Geophysical Research Letters}{48}{9}{e2021GL093152}.
\newblock
\begin{APACrefURL}
  \url{https://agupubs.onlinelibrary.wiley.com/doi/abs/10.1029/2021GL093152}
  \end{APACrefURL}
\newblock
\APACrefnote{e2021GL093152 2021GL093152}
\newblock
\begin{APACrefDOI} \doi{https://doi.org/10.1029/2021GL093152} \end{APACrefDOI}
\PrintBackRefs{\CurrentBib}

\bibitem [\protect \citeauthoryear {%
Skeltved%
, Ostgaard%
, Carlson%
, Gjesteland%
\BCBL {}\ \BBA {} Celestin%
}{%
Skeltved%
\ \protect \BOthers {.}}{%
{\protect \APACyear {2014}}%
}]{%
Skeltved2014}
\APACinsertmetastar {%
Skeltved2014}%
\begin{APACrefauthors}%
Skeltved, A.%
, Ostgaard, N.%
, Carlson, B.%
, Gjesteland, T.%
\BCBL {}\ \BBA {} Celestin, S.%
\end{APACrefauthors}%
\unskip\
\newblock
\APACrefYearMonthDay{2014}{11}{}.
\newblock
{\BBOQ}\APACrefatitle {Modelling the Relativistic Runaway Electron Avalanche
  and the feedback mechanism with {GEANT4}} {Modelling the relativistic runaway
  electron avalanche and the feedback mechanism with {GEANT4}}.{\BBCQ}
\newblock
\APACjournalVolNumPages{Journal of Geophysical Research: Space
  Physics}{119}{}{}.
\newblock
\begin{APACrefDOI} \doi{10.1002/2014JA020504} \end{APACrefDOI}
\PrintBackRefs{\CurrentBib}

\bibitem [\protect \citeauthoryear {%
Stadnichuk%
, Zelenyy%
, Nozik%
\BCBL {}\ \BBA {} Dolgonosov%
}{%
Stadnichuk%
\ \protect \BOthers {.}}{%
{\protect \APACyear {2019}}%
}]{%
Stadnichuk2019}
\APACinsertmetastar {%
Stadnichuk2019}%
\begin{APACrefauthors}%
Stadnichuk, E.%
, Zelenyy, M.%
, Nozik, A.%
\BCBL {}\ \BBA {} Dolgonosov, M.%
\end{APACrefauthors}%
\unskip\
\newblock
\APACrefYearMonthDay{2019}{}{}.
\newblock
{\BBOQ}\APACrefatitle {Monte Carlo simulation of the relativistic feedback
  discharge model (RFDM)} {Monte carlo simulation of the relativistic feedback
  discharge model (rfdm)}.{\BBCQ}
\newblock
\BIn{} (\BPG~164).
\newblock
\APACaddressPublisher{Armenia}{CRD Cosmic Ray Division, A Alikhanyan National
  Laboratory, Yerevan, Armenia}.
\newblock
\begin{APACrefURL}
  \url{http://inis.iaea.org/search/search.aspx?orig_q=RN:50069601}
  \end{APACrefURL}
\newblock
\APACrefnote{PHYSICS OF ELEMENTARY PARTICLES AND FIELDS}
\PrintBackRefs{\CurrentBib}

\bibitem [\protect \citeauthoryear {%
Stadnichuk%
, Zemlianskay%
\BCBL {}\ \BBA {} Efremova%
}{%
Stadnichuk%
\ \protect \BOthers {.}}{%
{\protect \APACyear {2021}}%
}]{%
EGU}
\APACinsertmetastar {%
EGU}%
\begin{APACrefauthors}%
Stadnichuk, E.%
, Zemlianskay, D.%
\BCBL {}\ \BBA {} Efremova, V.%
\end{APACrefauthors}%
\unskip\
\newblock
\APACrefYearMonthDay{2021}{}{}.
\newblock
{\BBOQ}\APACrefatitle {Simple "Reactor model" of relativistic runaway electron
  avalanches dynamics} {Simple "reactor model" of relativistic runaway electron
  avalanches dynamics}.{\BBCQ}
\newblock
\BIn{} \APACrefbtitle {EGU General Assembly 2021, online, 19–30 Apr 2021}
  {Egu general assembly 2021, online, 19–30 apr 2021}\ (\BVOL\ EGU21-13395).
\newblock
\begin{APACrefDOI} \doi{https://doi.org/10.5194/egusphere-egu21-13395}
  \end{APACrefDOI}
\PrintBackRefs{\CurrentBib}

\bibitem [\protect \citeauthoryear {%
Stolzenburg%
\ \BBA {} Marshall%
}{%
Stolzenburg%
\ \BBA {} Marshall%
}{%
{\protect \APACyear {2008}}%
}]{%
Stolzenburg2008}
\APACinsertmetastar {%
Stolzenburg2008}%
\begin{APACrefauthors}%
Stolzenburg, M.%
\BCBT {}\ \BBA {} Marshall, T.%
\end{APACrefauthors}%
\unskip\
\newblock
\APACrefYearMonthDay{2008}{07}{}.
\newblock
{\BBOQ}\APACrefatitle {Serial profiles of electrostatic potential in five {New
  Mexico} thunderstorms} {Serial profiles of electrostatic potential in five
  {New Mexico} thunderstorms}.{\BBCQ}
\newblock
\APACjournalVolNumPages{Journal of Geophysical Research}{113}{}{}.
\newblock
\begin{APACrefDOI} \doi{10.1029/2007JD009495} \end{APACrefDOI}
\PrintBackRefs{\CurrentBib}

\bibitem [\protect \citeauthoryear {%
Tessendorf%
, Rutledge%
\BCBL {}\ \BBA {} Wiens%
}{%
Tessendorf%
\ \protect \BOthers {.}}{%
{\protect \APACyear {2007}}%
}]{%
Tessendorf2007Multicell}
\APACinsertmetastar {%
Tessendorf2007Multicell}%
\begin{APACrefauthors}%
Tessendorf, S.%
, Rutledge, S.%
\BCBL {}\ \BBA {} Wiens, K.%
\end{APACrefauthors}%
\unskip\
\newblock
\APACrefYearMonthDay{2007}{11}{}.
\newblock
{\BBOQ}\APACrefatitle {Radar and Lightning Observations of Normal and Inverted
  Polarity Multicellular Storms from STEPS} {Radar and lightning observations
  of normal and inverted polarity multicellular storms from steps}.{\BBCQ}
\newblock
\APACjournalVolNumPages{Monthly Weather Review - MON WEATHER REV}{135}{}{}.
\newblock
\begin{APACrefDOI} \doi{10.1175/2007MWR1954.1} \end{APACrefDOI}
\PrintBackRefs{\CurrentBib}

\bibitem [\protect \citeauthoryear {%
Torii%
\ \protect \BOthers {.}}{%
Torii%
\ \protect \BOthers {.}}{%
{\protect \APACyear {2009}}%
}]{%
Torii2009}
\APACinsertmetastar {%
Torii2009}%
\begin{APACrefauthors}%
Torii, T.%
, Sugita, T.%
, Tanabe, S.%
, Kimura, Y.%
, Kamogawa, M.%
, Yajima, K.%
\BCBL {}\ \BBA {} Yasuda, H.%
\end{APACrefauthors}%
\unskip\
\newblock
\APACrefYearMonthDay{2009}{07}{}.
\newblock
{\BBOQ}\APACrefatitle {Gradual increase of energetic radiation associated with
  thunderstorm activity at the top of {Mt. Fuji}} {Gradual increase of
  energetic radiation associated with thunderstorm activity at the top of {Mt.
  Fuji}}.{\BBCQ}
\newblock
\APACjournalVolNumPages{Geophysical Research Letters}{36}{}{}.
\newblock
\begin{APACrefDOI} \doi{10.1029/2008GL037105} \end{APACrefDOI}
\PrintBackRefs{\CurrentBib}

\bibitem [\protect \citeauthoryear {%
Wada%
\ \protect \BOthers {.}}{%
Wada%
\ \protect \BOthers {.}}{%
{\protect \APACyear {2019}}%
}]{%
Wada2019}
\APACinsertmetastar {%
Wada2019}%
\begin{APACrefauthors}%
Wada, Y.%
, Enoto, T.%
, Nakamura, Y.%
, Furuta, Y.%
, Yuasa, T.%
, Nakazawa, K.%
\BDBL {}Tsuchiya, H.%
\end{APACrefauthors}%
\unskip\
\newblock
\APACrefYearMonthDay{2019}{06}{}.
\newblock
{\BBOQ}\APACrefatitle {Gamma-ray glow preceding downward terrestrial gamma-ray
  flash} {Gamma-ray glow preceding downward terrestrial gamma-ray
  flash}.{\BBCQ}
\newblock
\APACjournalVolNumPages{Communications Physics}{2}{}{67}.
\newblock
\begin{APACrefDOI} \doi{10.1038/s42005-019-0168-y} \end{APACrefDOI}
\PrintBackRefs{\CurrentBib}

\bibitem [\protect \citeauthoryear {%
Williams%
}{%
Williams%
}{%
{\protect \APACyear {1989}}%
}]{%
Williams1989}
\APACinsertmetastar {%
Williams1989}%
\begin{APACrefauthors}%
Williams, E.%
\end{APACrefauthors}%
\unskip\
\newblock
\APACrefYearMonthDay{1989}{01}{}.
\newblock
{\BBOQ}\APACrefatitle {The tripole structure of thunderstorm} {The tripole
  structure of thunderstorm}.{\BBCQ}
\newblock
\APACjournalVolNumPages{Journal of Geophysical Research}{941}{}{13151-13167}.
\newblock
\begin{APACrefDOI} \doi{10.1029/JD094iD11p13151} \end{APACrefDOI}
\PrintBackRefs{\CurrentBib}

\bibitem [\protect \citeauthoryear {%
Zelenyi%
, Nozik%
\BCBL {}\ \BBA {} Stadnichuk%
}{%
Zelenyi%
\ \protect \BOthers {.}}{%
{\protect \APACyear {2019}}%
}]{%
doi:10.1063/1.5130111}
\APACinsertmetastar {%
doi:10.1063/1.5130111}%
\begin{APACrefauthors}%
Zelenyi, M.%
, Nozik, A.%
\BCBL {}\ \BBA {} Stadnichuk, E.%
\end{APACrefauthors}%
\unskip\
\newblock
\APACrefYearMonthDay{2019}{}{}.
\newblock
{\BBOQ}\APACrefatitle {Reactor like {TGE} model} {Reactor like {TGE}
  model}.{\BBCQ}
\newblock
\APACjournalVolNumPages{{AIP} Conference Proceedings}{2163}{1}{060005}.
\newblock
\begin{APACrefDOI} \doi{10.1063/1.5130111} \end{APACrefDOI}
\PrintBackRefs{\CurrentBib}

\bibitem [\protect \citeauthoryear {%
{Zelenyi, Mikhail}%
, {Stadnichuk, Egor}%
\BCBL {}\ \BBA {} {Nozik, Alexander}%
}{%
{Zelenyi, Mikhail}%
\ \protect \BOthers {.}}{%
{\protect \APACyear {2019}}%
}]{%
refId0}
\APACinsertmetastar {%
refId0}%
\begin{APACrefauthors}%
{Zelenyi, Mikhail}%
, {Stadnichuk, Egor}%
\BCBL {}\ \BBA {} {Nozik, Alexander}.%
\end{APACrefauthors}%
\unskip\
\newblock
\APACrefYearMonthDay{2019}{}{}.
\newblock
{\BBOQ}\APACrefatitle {Calculation of gain coefficient in {Dwyer} relativistic
  discharge feedback model of thunderstorm runway breakdown} {Calculation of
  gain coefficient in {Dwyer} relativistic discharge feedback model of
  thunderstorm runway breakdown}.{\BBCQ}
\newblock
\APACjournalVolNumPages{EPJ Web Conf.}{201}{}{07003}.
\newblock
\begin{APACrefURL} \url{https://doi.org/10.1051/epjconf/201920107003}
  \end{APACrefURL}
\newblock
\begin{APACrefDOI} \doi{10.1051/epjconf/201920107003} \end{APACrefDOI}
\PrintBackRefs{\CurrentBib}

\bibitem [\protect \citeauthoryear {%
Østgaard%
\ \protect \BOthers {.}}{%
Østgaard%
\ \protect \BOthers {.}}{%
{\protect \APACyear {2019}}%
}]{%
10_month_ASIM}
\APACinsertmetastar {%
10_month_ASIM}%
\begin{APACrefauthors}%
Østgaard, N.%
, Neubert, T.%
, Reglero, V.%
, Ullaland, K.%
, Yang, S.%
, Genov, G.%
\BDBL {}Al-nussirat, S.%
\end{APACrefauthors}%
\unskip\
\newblock
\APACrefYearMonthDay{2019}{}{}.
\newblock
{\BBOQ}\APACrefatitle {First 10 Months of TGF Observations by ASIM} {First 10
  months of tgf observations by asim}.{\BBCQ}
\newblock
\APACjournalVolNumPages{Journal of Geophysical Research:
  Atmospheres}{124}{24}{14024-14036}.
\newblock
\begin{APACrefURL}
  \url{https://agupubs.onlinelibrary.wiley.com/doi/abs/10.1029/2019JD031214}
  \end{APACrefURL}
\newblock
\begin{APACrefDOI} \doi{https://doi.org/10.1029/2019JD031214} \end{APACrefDOI}
\PrintBackRefs{\CurrentBib}

\end{thebibliography}

%
%
%
%
%

\section{Appendix A: Gamma-radiation dynamics in RREA models with positive feedback}
\label{appendix_a}

Let us consider the dynamics of RREAs with relativistic feedback and external source of seed particles. We aim to calculate the dependence of the particle flux on time for the cell of length L with the flux of external seed electrons $I_{SE}$. The feedback coefficient $\gamma$ is the average number of RREAs produced by one avalanche. The increase in the number of avalanches in case of zero flux of seed particles is provided only by feedback \cite{Dwyer_2003_fundamental_limit}:

\begin{equation}
    d N_{RREA} = N_{RREA}(0) \cdot (\gamma - 1) \cdot \frac{c}{2 L} dt
    \label{appendix_A_dif}
\end{equation}

where $N_{RREA}(0)$ is the number of runaway electron avalanches in the cell at the initial moment. $\frac{2L}{c}$ is the duration of one feedback cycle, equal to twice the time of photon propagation through the cell. The factor $(\gamma - 1)$ means that $(\gamma - 1)$ new RREAs are born in one cycle of feedback. If $\gamma = 1$, then the avalanches are self-sustaining: $N_{RREA} = const$. The solution to the equation \ref{appendix_A_dif} is:

\begin{equation}
    N_{RREA} (t) = N_{RREA}(0) \cdot e^{\frac{c}{2 L} (\gamma - 1) t}
\end{equation}

For $\gamma = 1$ all RREAs developed in the cell remain there. Consequently, in \add{the} case of the flux of seed electrons $I_{SE}$, the accumulation of avalanches occurs as follows:

\begin{equation}
    d N_{RREA} = I_{SE} \cdot S \cdot dt
\end{equation}

where $S$ is the area of the cell perpendicular to the field direction. Thus, under the considered conditions, the number of avalanches grows linearly with time:

\begin{equation}
    N_{RREA} (t) = I_{SE} \cdot S \cdot t
\end{equation}

In the presence of feedback and seed particles the number of avalanches takes the following form:

\begin{equation}
    d N_{RREA} = N_{RREA} \cdot (\gamma - 1) \cdot \frac{c}{2 L} dt + I_{SE} \cdot S \cdot dt
\end{equation}

By replacing $\alpha = N_{RREA} + \frac{2L}{c\left(\gamma-1\right)} S I_{SE}$, the equation is reduced to an equation with separable variables, the solution of which with initial condition $N_{RREA}(0) = 0$ is as follows:

\begin{equation}
    N_{RREA} (t) = N_{RREA}(0) e^{\frac{c}{2 L} (\gamma - 1) t} + I_{SE} S \frac{2 L}{(\gamma - 1) c} (e^{\frac{c}{2 L} (\gamma - 1) t} - 1)
    \label{rreas_dynamics}
\end{equation}

There is an alternative approach to the same problem. Let $I_{SE}Sd\tau$ of seed particles arrive at the cell at the moment $\tau$. Then by the time t they will multiply due to relativistic feedback, and their number will become equal to $dN_{RREA} = I_{SE} S e^{\frac{c}{2L} (\gamma - 1) (t - \tau)} d\tau$. Integration of this expression over $\tau$ leads to the Formula~(\ref{rreas_dynamics}) describing the number of avalanches in a cell. Provided that one RREA produce $N_{particles~from~RREA}$ particles (for example, high energy photons or positrons) during one feedback cycle, the total number of particles of these type depends on time as follows:

\begin{equation}
    N_{particles~total} (t) = N_{particles~from~RREA} \cdot N_{RREA} (t)
\end{equation}

The same formalism can be applied to the analytical description of the multicell reactor model. If we define the global multiplication factor $\varepsilon = \gamma - 1$ then the increase in the concentration of high energy photons in the point $(r, z)$ at the moment t is following:

\begin{equation}
    dn(r, z, t) = \frac{\partial n_{cosmic}}{\partial t} dt + n(r, z, t) \cdot \varepsilon dt
\label{eq:dif_cosmic}
\end{equation}

The solution of (\ref{eq:dif_cosmic}) satisfying the initial condition $n(r, z, 0) \equiv n_0$ is:

\begin{equation}
    n(r, z, t) = n_0e^{\varepsilon t} + \frac{\frac{\partial n_{cosmic}}{\partial t}}{\varepsilon} \cdot \bigg(e^{\varepsilon t} - 1\bigg)
\end{equation}

\begin{figure}[h!]
    \centering
    \includegraphics[width=0.85\textwidth]{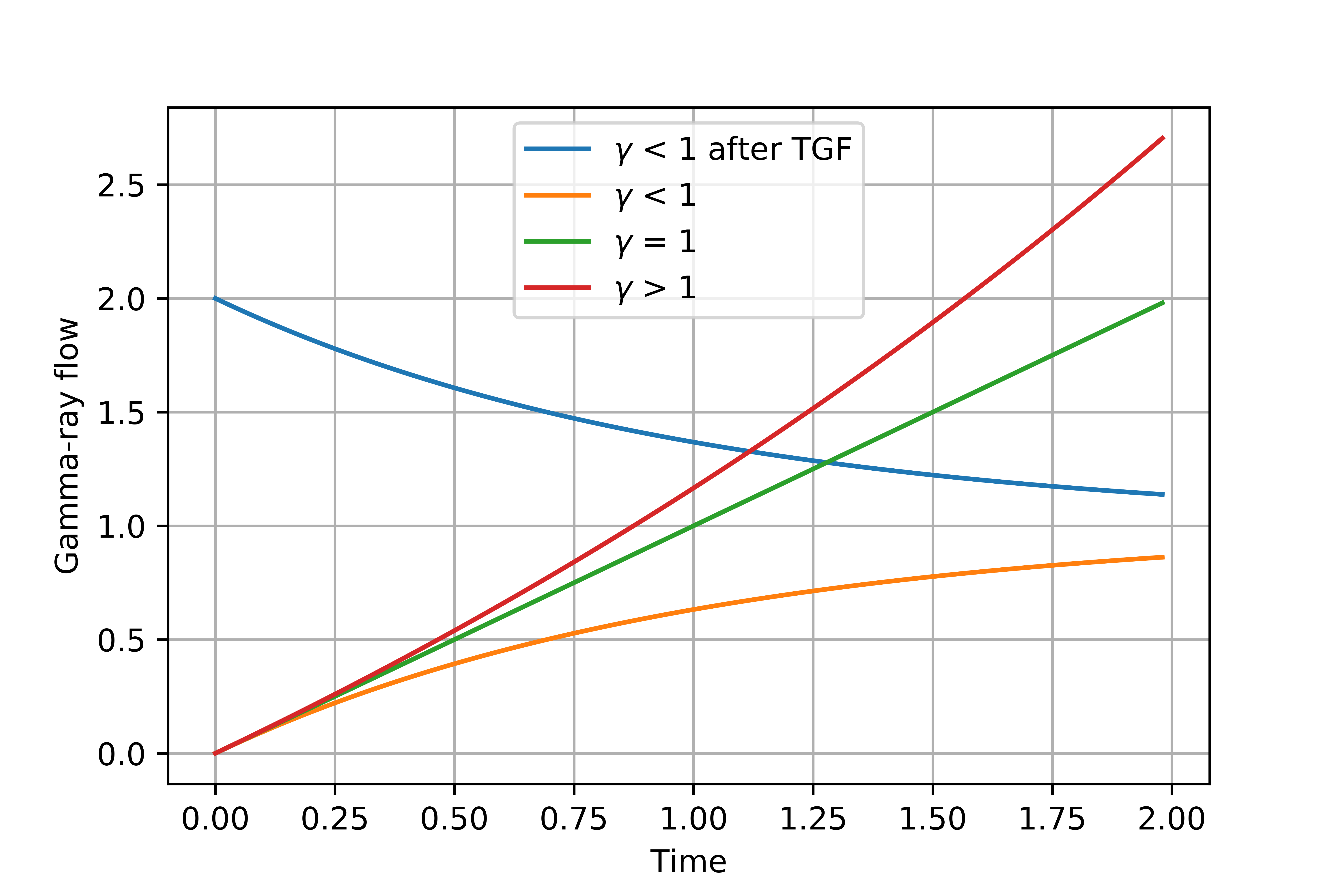}
    \caption{Dynamics of the gamma-ray flux within a RREA model with positive feedback in the presence of an external constant source of seed particles, Formula~\ref{rreas_dynamics}. $\gamma$ is the average ratio of the number of particles in the next generation of feedback to the number of particles in the previous generation. The exponential increase of the flux corresponds to $\gamma > 1$. In the case of $\gamma$ = 1 the flux grows linearly as the thunderstorm acts as the avalanche accumulator. For $(\gamma < 1)$ the gamma-ray flow asymptotically tends to a constant flow, which is higher relative to RREAs radiation without feedback processes by a factor $\frac{1}{1 - \gamma}$.
    }
    \label{feedback_dynamics}
\end{figure}

The presented consideration leads to the following conclusion, common for all models with positive feedback in the dynamics of RREAs. In the case of $\gamma = 1$ the flux grows linearly, the avalanches are self-sustaining. The linear increase of the number of RREAs $N_{RREA} (t) = I_{SE} S t$ can be obtained from Formula~(\ref{rreas_dynamics}) by Taylor expansion in the small parameter $(\gamma - 1)$. Therefore, even at $\gamma$ = 1, a TGF-like event can be generated by the feedback mechanism. For $\gamma > 1$ gamma-ray flux increases exponentially in time. $\gamma < 1$ leads to the exponential decay of the flux with the asymptotic constant value, which is higher than RREAs radiation without feedback by a factor $\frac{1}{1 - \gamma}$. For $(\gamma < 1)$ the factor $\left( e^{\frac{c}{2L}\left(\gamma-1\right)t}-1 \right)$ decreases in time (this factor is negative, and $(\gamma - 1)$ in the denominator of Formula~(\ref{rreas_dynamics}) is also negative, therefore, the total number of avalanches is positive). The resulting dynamics of the number of RREAs is an exponential growth gradually turning into a constant value. The greater the $\gamma$, the greater the final constant flux. Thus, strong feedback is not required to describe gamma-ray glows and TGE. Finally, if the initial gamma-ray flux is high, for example, just after TGF peak, and $(\gamma < 1)$, then the flux will decay exponentially. This fact might explain TGF decays within the framework of models of RREAs dynamics with feedback (relativistic feedback discharge model or reactor model), Figure~\ref{feedback_dynamics}.

\section{Appendix B: Microscopic simulation comments}
The modeling of RREA evolution is carried out using GEANT4 toolkit (physics list: G4EmStandartPhysics$\_$option4) in two stages. In the first stage, an attenuation length of a gamma-ray photon $\lambda_{\gamma}$ is calculated. A rectangular volume with air is modeled, at the end of which the detector is located. A 7~MeV gamma-ray is launched in the direction of the detector. With the increase of the distance to the detector number of gamma-rays reaching it decreases. Number of gamma rays reaching the detector depending on the distance to the detector is fitted with $N_{\gamma}(z) = N_0 e^{-\frac{z}{\lambda_{\gamma}}}$ ($N_0$ is number of gamma-rays launched in the simulation). In this way, the attenuation length of the gamma-ray photon is found. The results of modeling for different values of the air density are presented in Figure~\ref{gamma_decay}. The second stage of modeling provides information on the characteristic path length of the gamma-ray for the generation of a runaway electron. A gamma-ray photon is launched in a rectangular air cell with a 4~km length. As the particle moves in the cell, secondary particles are generated. Information on secondary electrons is registered at the moment of birth and then electrons are stopped to get rid of their influence on the simulation results. After receiving information about the created electrons, one can filter out particles with subcritical energy for the corresponding electric field strength. The filtered data may be approximated as follows: $dN_{e^{-}}(z) = N_{\gamma}(0)e^{-\frac{z}{\lambda_{\gamma}}}\frac{dz}{\lambda_{e^{-}}}$, $\lambda_{\gamma}$ ---- gamma attenuation length, $N_{\gamma}(0)$ --- initial number of gamma,
$\lambda_{e^{-}}$ --- mean free path of gamma governing the production of a runaway electron, $N_{e^{-}}(z)$ --- the number of electrons with the energy above the runaway threshold.

\section{Appendix C: Runaway electron propagation between thunderstorm critical electric field regions}

\label{Appendix_D}

More accurate consideration of runaway electron transport between thunderstorm critical electric field regions (cells) requires taking into account the distance between cells, which in general does not contain the electric field sufficient for the electrons to run away. In this section, a brief analysis of the effect of the distance between cells on runaway electrons exchange between cells is presented.

Let all RREAs accelerated by cells be fully developed, with a fully developed spectrum. Let the average distance between cells be equal $l$. Let the average subcritical electric field between cells be $E_{out}$. Let the value of the critical electric field at the considered height be equal $E_{c}$. Therefore, the energy that an electron loses when passing between the cells can be estimated as follows:

\begin{equation}
    \varepsilon_{out} = e l \left( E_c - E_{out} \right)
\end{equation}

Let the maximum energy of runaway electrons be equal $\varepsilon_{max}$, it can be estimated as 40~MeV \cite{asim_spectrum}. Let the critical energy (minimum runaway electron energy) be equal $\varepsilon_{min}$. Thus, knowing the spectrum of runaway electrons, the fraction of surviving electrons in transit between cells can be estimated as follows:

\begin{equation}
    \delta = \frac{\int_{\varepsilon_{out} + \varepsilon_{min}}^{\varepsilon_{max}}d\varepsilon f(\varepsilon)}{\int_{\varepsilon_{min}}^{\varepsilon_{max}}d\varepsilon f(\varepsilon)} S\left( \varepsilon_{max} - \varepsilon_{out} - \varepsilon_{min} \right)
\end{equation}

Taking into account the generally accepted spectrum of runaway electrons \cite{Babich_2020}, the fraction of propagated electrons will be as follows:

\begin{equation}
    \delta = \frac{\int_{\varepsilon_{out} + \varepsilon_{min}}^{\varepsilon_{max}} d\varepsilon e^{-\frac{\varepsilon}{7.3~MeV}}}{\int_{\varepsilon_{min}}^{\varepsilon_{max}}d\varepsilon e^{-\frac{\varepsilon}{7.3~MeV}}}S\left( \varepsilon_{max} - \varepsilon_{out} - \varepsilon_{min} \right)
\end{equation}

Consequently,

\begin{equation}
    \delta = \frac{e^{-\frac{\varepsilon_{out} + \varepsilon_{c}}{7.3~MeV}} - e^{-\frac{\varepsilon_{max}}{7.3~MeV}}}{e^{-\frac{\varepsilon_{c}}{7.3~MeV}} - e^{-\frac{\varepsilon_{max}}{7.3~MeV}}}S\left( \varepsilon_{max} - \varepsilon_{out} - \varepsilon_{c} \right)
\end{equation}

This leads to the following local multiplication factor taking into account the electron transport, Formula~\ref{formula_nu_rreas}:

\begin{equation}
\begin{split}
    \nu = \frac{P}{L} \frac{\lambda_{RREA}}{\lambda_{e^- \rightarrow \gamma}}\left(\lambda_{RREA}e^{\frac{l}{\lambda_{RREA}}} - \lambda_{RREA} - L\right) + \\
    + \frac{\lambda_{RREA}}{\lambda_{e^- \rightarrow \gamma}}\left(e^{\frac{L}{\lambda_{RREA}}} - 1\right) \cdot \frac{0.5\delta P \frac{\lambda_{RREA}}{L} \left(e^{\frac{L}{\lambda_{RREA}}} - 1\right) }{1 - 0.5\delta e^{\frac{L}{\lambda_{RREA}}}}
\end{split}
    \label{electron_decay}
\end{equation}

\begin{figure}
    \centering
    \includegraphics[width=0.85\textwidth]{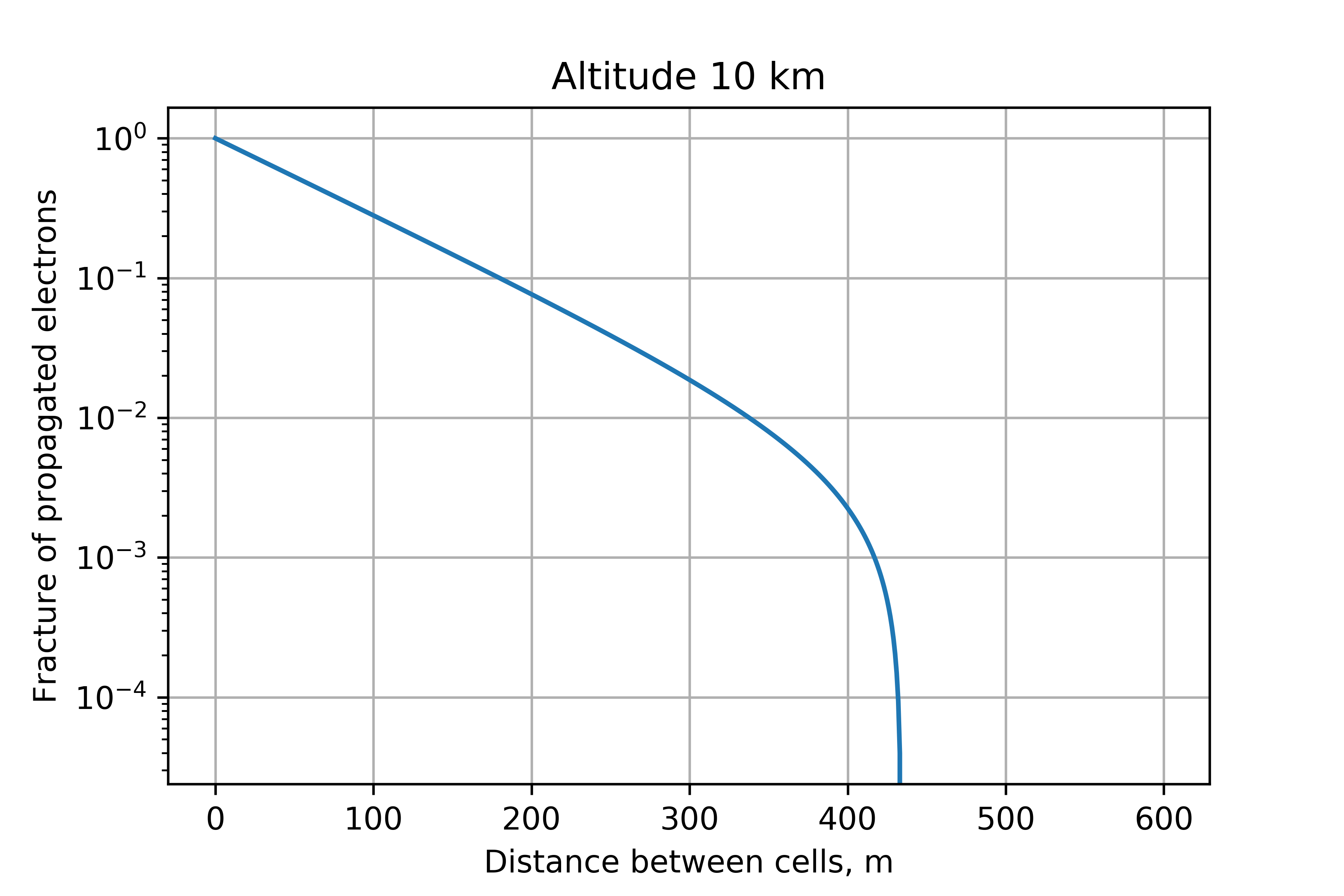}
    \caption{Fraction of runaway electrons reaching other thunderstorm runaway electron accelerating regions in dependence on the distance between cells, Formula~\ref{electron_decay}. Here the electric field strength between cells is considered to be zero. The graph shows that electron transport between cells should be considered in the reactor feedback if the distance between cells is less than 200~m (for 10~km altitude).}
    \label{fig:fraction_electron_transport}
\end{figure}

The estimation (Formula~\ref{electron_decay}) shows that for 10~km altitude runaway electron transport between cells should be taken into account if the distance between cells is below approximately 200~m, Figure~\ref{fig:fraction_electron_transport}. In other cases, runaway electron transport is negligible for reactor feedback models.

\end{document}